\RequirePackage{fix-cm}

\documentclass[smallextended,runningheads,natbib]{svjour3}       % onecolumn (second format)
\smartqed
\makeatletter
\let\cl@chapter\relax
\makeatother
\usepackage{amssymb,amsmath,amsfonts}
\usepackage{graphicx,psfrag,color}
\usepackage{tikz}
\usepackage{mathrsfs}
\usepackage{ragged2e}
\usepackage{tabularx}
\usepackage{mathtools}

\def\url#1{\expandafter\string\csname #1\endcsname}

\usepackage[left=1.0in, right=1.0in, top=1.0in, bottom=1.0in,
  includehead,includefoot]{geometry}

\graphicspath{{./include/}{./include/msc_figs/}{./include/constvol/}}
\makeatletter
\def\input@path{{./include/}{./include/msc_figs/}}
\makeatother

\usepackage[T1]{fontenc}
\usepackage[utf8]{inputenc}
\usepackage{lmodern}
\usepackage[english]{babel}
\usepackage{csquotes}
\usepackage[capitalise]{cleveref}
\usepackage[multidot]{grffile} % allow having dots in graphics filenames

\usepackage{multirow}
\usepackage{booktabs}
\usepackage{rotating}

\usepackage{xspace}
\usepackage{mathtools,xparse}

\DeclarePairedDelimiterXPP\seq[1]{}{(}{)}{}{#1}

\newcommand{\dd}{\ensuremath{\, \mathrm{d}}}

\newcommand{\sgn}{\operatorname{sgn}}

\newcommand{\lb}{\left(}
\newcommand{\rb}{\right)}

\newcommand{\lsb}{\left[\, }
\newcommand{\rsb}{\,\right] }

\usepackage{siunitx}
\DeclareSIUnit\molar{\mole\per\cubic\deci\metre}
\DeclareSIUnit\Molar{\textsc{m}}

\newcolumntype{b}{X}
\newcolumntype{Y}{>{\RaggedRight\arraybackslash}X}

\newtheorem{result}{Result}
\crefname{result}{Result}{Results}
\Crefname{result}{Result}{Results}
\journalname{}

\begin{document}

%\setcounter{tocdepth}{4}
%\setcounter{secnumdepth}{3}
%\tableofcontents

\title{Cell size, mechanical tension, and GTPase signaling in the Single Cell}

% \subtitle{}

\author{Andreas Buttenschön \and
        Yue Liu \and
        Leah Edelstein-Keshet
}


\institute{A.\ Buttenschön \at
          Department of Mathematics, University of British Columbia, Vancouver
          V6T~1Z2, BC, Canada\\
          \email{abuttens@math.ubc.ca}           %  \\
          \and
          Y.\ Liu \at
          Department of Mathematics, University of British Columbia, Vancouver
          V6T~1Z2, BC, Canada\\
          \email{liuyue@math.ubc.ca}
          \and
          L.\ Edelstein-Keshet \at
          Department of Mathematics, University of British Columbia, Vancouver
          V6T~1Z2, BC, Canada\\
          \email{keshet@math.ubc.ca}           %  \\
}

\date{Received: date / Accepted: date}

\maketitle

\begin{abstract}
Cell polarization requires redistribution of specific proteins to the nascent
front and back of a eukarytotic cell. Among these proteins are Rac and Rho,
members of the small GTPase family that regulate the actin cytoskeleton. Rac
promotes actin assembly and  protrusion of the front edge, whereas Rho
activates myosin-driven contraction at the back.  Mathematical models of
cell polarization at many levels of detail have appeared. One of the
simplest based on ``wave-pinning'', consists of a pair of reaction-diffusion
equations for a single GTPase. Mathematical analysis of wave-pinning so far
is largely restricted to static domains in one spatial dimension. Here we
extend the analysis to  cells that change in size, showing that both
shrinking and growing cells can lose polarity. We further consider the
feedback between mechanical tension, GTPase activation, and cell deformation
in both static, growing, shrinking, and moving cells. Special cases
(spatially uniform cell chemistry, absence or presence of mechanical
feedback) are analyzed, and the full model is explored by simulations in 1D.
We find a variety of novel behaviors, including ``dilution-induced''
oscillations of Rac activity and cell size, as well as gain or loss of
polarization and motility in the model cell.

\keywords{Cell size, motility, GTPases, Rac and Rho, cell polarization,
 reaction-diffusion equations, growing 1D domain}

\end{abstract}

\begin{acknowledgements}
LEK and coauthored are (partially) supported by NSERC (Natural Sciences and
Engineering Research Council) Discovery grant 41870 to LEK. YL was partially
supported by an NSERC postgraduate fellowship. AB was partially supported by an
NSERC PDF Fellowship.  We are grateful to the Pacific Institute for Mathematical
Sciences for providing space and resources for AB’s postdoctoral
research.  The authors thank C.\ Zmurchok and W.\ R.\ Holmes and E. \ Rens for
valuable discussions.
\end{acknowledgements}

%%%%%%%%%%%%%%%%%%%%%%%%%%%%%%%%%%%%%%%%%%%%%%%%%%%%%%%%%

\section{Introduction}

Cell polarization and motility play a vital role in living organisms across all
size scales. In normal development, wound healing, and immune surveillance,
cells undergo directed migration tightly controlled by intracellular signaling
pathways. Hence, these cellular phenomena have
attracted attention of modelers, and gained representation at a whole spectrum
of levels of detail, in 1D \citep{otsuji2007,Mori2008,Dawes2007,Holmes2012}, 1.5D
(the edge of a 2D domain) \citep{Meinhardt1999,neilson2011}, as well as 2D
\citep{Maree2006,wolgemuth2010,Maree2012,Camley2013}, and 3D \citep{Cusseddu2018}
geometries.

An elementary model for cell polarization by ``wave-pinning'' was introduced in
\citep{Mori2008} in 1D, based on a single signaling protein (GTPase) inside a
stationary cell of fixed length. The wave-pinning model was a simplification of
the more biologically detailed models of GTPase role in polarity captured in
\citep{Maree2012,Jilkine2011}. Generalization of this model to a pair of mutually
antagonistic (Rac-Rho) GTPases and use of approximation methods demonstrated
coexistance of cell states such as polarized, low uniform and high uniform
protein activity levels \citep{holmes2016}.

As interesting aspect of the GTPases Rac and Rho is that they tend to
concentrate at cell edges, where they drive actin assembly and edge protrusion
(Rac) or actomyosin contraction (Rho), which moves the cell edge outwards or
inwards. The cell tends to resist deformation, which means that when one edge
moves, the opposite edge is pulled or pushed accordingly. At the same time,
protrusion or retraction may cause cell volume to change, particularly in
combination with aquaporin channels that allow water to flow in or out of the
cell. Hence, cells can undergo fluctuations in size \citep{Zehnder2015} that
potentially affect their internal concentration.

In \citep{Mori2008,mori2011,Jilkine2011,holmes2016}, cell
size was assumed to be static, whereas here we lift this restriction.  The
effect of cell size was briefly considered in modeling the polarization response
of HeLa cells to Rac stimulation in
microfluidic channels \citep{Holmes2012}. It is also incorporated in a companion
paper by \citep{Zmurchok2019}, where the focus is on cell shape diversity arising
from autoactivation as well as mutual inhibition in Rac-Rho interactions. Cells
were deformable in \citep{Zmurchok2018}, but possibility dilution was neglected,
whereas here we consider such effect. Finally, in \citep{Zmurchok2018,Bui2019},
feedback from tension due to cell deformation was included, but only in model
cells with spatially uniform GTPase activity. Here we generalize to full
spatially distributed GTPase activity, which introduces new interesting
behavior.

Mathematical analysis is possible for simple geometries and models with few
components, and becomes increasingly forbidding as the size of the signaling
network or the complexity of the geometry increases. For this reason, we
concentrate on a toy model in 1D geometry to describe the interplay between cell
size, cell polarization/motility and mechanical feedback. We show that this
interplay can lead to cycles of expansion and contraction in certain cases
and/or initiation of motility after some delay time.

The model is simple enough that some analytical results can be written to
characterize regimes of behaviour. We can apply several methods,  including
linear stability analysis, approximations of kinetic terms by piecewise linear
functions (where steady states can be found in closed form), as well as
bifurcation analysis and numerical PDE solutions. Using these methods, we here
generalize and extend some of the previous mathematical findings about
wave-pinning, and help to increase insight about how such mechanisms for cell
polarization could operate.

\begin{figure}\centering
    \includegraphics[width=0.6\textwidth]{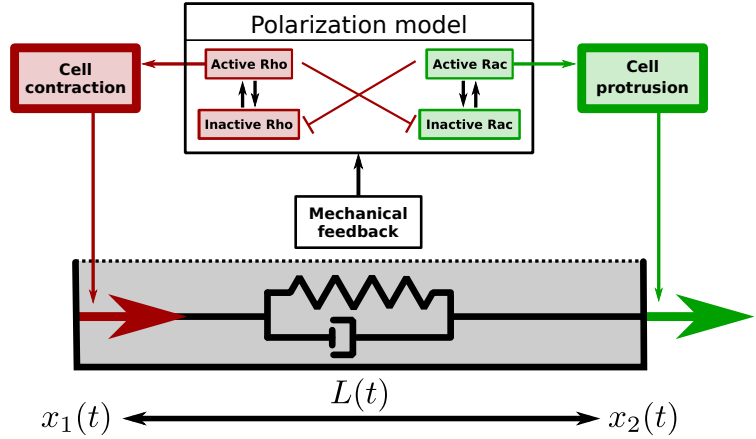}
    \caption{The model considered here has two key modules: the polarization
    model, and the one-dimensional visco-elastic cell model. Here a Kelvin-Voigt
    element is placed between the cell endpoints ($x_1(t)$ and $x_2(t)$). While
    other visco-elastic models are possible, here we focus on the interplay
    between mechanics and signalling. The polarization model captures the
    spatio-temporal dynamics of one or more small GTPases affecting the cell's
    cytoskeleton. Here their effects i.e.\ assembly of cell protrusions (Rac) or
    cell contractions (Rho) are captured by applying forces on the cell's
    endpoints. Here we consider three different signalling modules.
    For a spatially uniform single GTPase module (either Rac or Rho) we present
    detailed phase-plane and bifurcation analysis in \cref{sec:uniform_results}.
    In particular, we demonstrate the existence of ``dilution-induced''
    oscillations of Rac concentration and cell size. In \cref{sec:spatial_results}
    we show that a spatially heterogenous GTPase (Rac or Rho) allows the cell to
    polarize and move. However, if a cell is to soft i.e.\ it undergoes large
    size changes, it may undergo ``dilution-induced'' depolarize. Finally, in
    \cref{sec:RacRho_results} we consider a polarization module consiting of
    antagonistic Rac and Rho. We capture all features of the single GTPase
    models, but also gain more complicated spatio-temporal patterns.
}\label{Fig:ModelOverview}
\end{figure}

\section{Preexisting model to be generalized}
\label{sec:Modeleqs}
We briefly outline previous work \citep{Mori2008,holmes2016} as a starting point
for our paper. Concentrations of active ($G$) and inactive ($G_i$) GTPases are
modeled by reaction-diffusion (RD) equations,
\begin{equation}\label{Eqn:General_GTP_model}
\begin{split}
 G_t     &= D   \Delta G + A(G) G_i - I G,\\ %\quad x \in \Gamma(t), t \ge 0
 (G_i)_t &= D_i \Delta G_i - A(G) G_i + I G. %\quad x \in \Gamma(t), t \ge 0. \\
\end{split}
\end{equation}

Active GTPase is membrane-bound, so the rates of diffusion satisfy $D \ll D_i$.
$A(G)$ is a rate of activation and $I$ a rate of inactivation. Feedback from the
active GTPase to its kinetics is assumed to be
directed through the rate of activation of GTPase by GEFs (guanine nucleotide
exchange factors), whereas the rate of inactivation by GAPs (GTPase activating
proteins) is taken as constant. Since the GTPase cannot enter or leave through
the cell edge, the PDEs are supplemented by Neumann (no-flux) boundary
conditions. Equations~\eqref{Eqn:General_GTP_model} conserve the total amount
of GTPase ($G_T$) in a cell, that is
\[
 G_T \coloneqq \int \lb G + G_i \rb \dd x.
\]
As in previous papers, we consider both the single GTPase (``wave-pinning'') model of \citep{Mori2008}, as well as the Rac-Rho model of \citep{holmes2016}.
%The reaction-diffusion equations~\eqref{Eqn:General_GTP_model}, have interesting behavior when their kinetic terms lead to bistability. This is possible for either positive feedback or double negative feedback, leading to two cases we consider:

\begin{description}
    \item[\bf Single GTPase model:]
        The standard functional form for the rate of activation, $A(G)$, is
        \[
            A(G) = \left( \beta + \gamma \frac{G^n}{1 + G^n}\right).
        \]
    \item[\bf The Rac-Rho (two GTPase) model:]
        Let $G_R, G_\rho$ represent the  activities of Rac and Rho GTPases, respectively.
        We assume mutual inhibition. That is, the rate of activation of Rac is reduced by Rho and vice versa.
        The standard function form for the activation rate of Rac is
        \[
            A_R(G_\rho) = \left( \beta + \gamma \frac{1}{1 + G_\rho^n}\right).
        \]
and similarly for the dependence
$A_\rho(G_R) $ of Rho activation rate on the activity of Rac \citep{holmes2016}.
\end{description}

%As in \citep{holmes2016}, these activation rate functions can be approximated by piecewise constant switches (the case of $n \to \infty$) where steady states can be written in explicit form. We describe such analysis further on.

%%%%%%%%%%%%%%%%%%%%%%%%%%%%%%%
\section{Model adaptation (1): Dynamic cell size}
\label{sec:ModelAdapt1}

\begin{figure}
    \centering
    \includegraphics[scale=0.4]{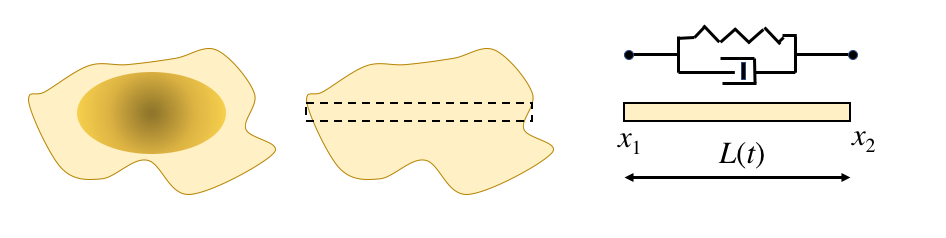}
    \caption{Left: Top-down view of a eukaryotic cell. Middle: the nucleus has
    been removed, so that the cell fragment is a thin sheet with approximately
    uniform thickness ($<1 \mu$m). A 1D transect across the cell is shown.
    Right: the front and rear positions of the cell edge are labeled. We
    consider cell motion such that $x_1, x_2, L$ are all possibly
    time-dependent. The region $x_1 \le x \le x_2$ is the domain inside which
    GTPase reaction-diffusion takes place. Mechanically, the 1D cell is
    represented as a spring dashpot (Kelvin-Voigt) element.}
    \label{fig:modelGeometry}
\end{figure}

The geometry for the single-cell model in \cref{fig:modelGeometry} represents a
transect across the diameter of the cell.

\begin{figure}
    \centering
    \includegraphics[scale=0.7]{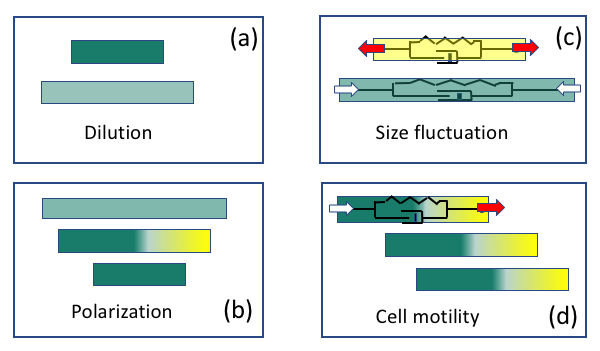}
    \caption{A summary of the main results of this paper. We discuss how changes
    in cell size can result in dilution of internal chemical concentration or
    activity (a). We show that size fluctuations can lead to loss (or gain) of
    chemical polarization by wave-pinning, as in \citep{Mori2008} (b). We then
    introduce mechanical force of protrusion (Rac) or contraction (Rho). In the
    uniform Rac case (c), cell size fluctuations can occur by virtue of dilution
    (and loss of Rac activity) when the cell expands. Finally, we show how the
    mechanochemical model linking GTPase activity to cell protrusion can lead to
    cell motility.}
    \label{fig:ResultsSchematic}
\end{figure}

We keep track of the coordinates for the front and rear cell edges $x_1(t),
x_2(t)$, where (WLOG) $x_1(t) < x_2(t)$, so that the cell's length is
$L(t) = x_2(t) - x_1(t) > 0$.

Denote by $\Gamma(t)$ the cell interior, i.e.\ the
%The interior of the cell is captured by the
interval $x_1 \le x \le x_2$.
%, which we denote $\Gamma(t)$.
%The activity of proteins that regulate cell motility (GTPases) will be described by reaction-diffusion equations on the domain $\Gamma(t)$.
A novel feature here is that reaction and diffusion of GTPases is considered on
this domain, whose size can change over time.

The motion of cell edges (and the domain $\Gamma(t)$) depend on the regulatory
proteins inside the cell. We consider several cases:

\begin{enumerate}
    \item A cell with one GTPase which causes edge protrusion (e.g.\ Rac).
    \item A cell with one GTPase which leads to edge contraction (e.g.\ Rho).
    \item A cell with both Rac and Rho, with mutual inhibition.
\end{enumerate}

The fact that the domain $\Gamma(t)$ can change with time requires that we
modify the RD equations \eqref{Eqn:General_GTP_model}. We address this in the
next section.

%We discuss both the spatially uniform variants of these GTPase models, as in \citep{holmes2016,Zmurchok2018}, as well as the full spatially distributed variants, where the chemical activity can become polarized. See also \citep{Zmurchok2019} where additional feedbacks are considered.

%%%%%%%%%%%%%%%%
%\subsection{The effect of domain expansion and contraction on the reaction-diffusion PDEs}
\subsection{Changing domain size affects the reaction-diffusion PDEs}

%The distribution of GTPases in a static cell domain is described by the PDEs \eqref{Eqn:General_GTP_model}.  But GTPases affect the cell size and shape.

In principle, when cell edges protrude or retract, the size of the membrane and
cytosol compartments change. Cell volume is known to fluctuate in experimental
settings, e.g.\ when water enters or leaves the cell through aquaporin channels
\citep{Zehnder2015}. Here we consider the simplest assumption, that membrane and
cytosol compartment sizes both vary proportionately as the cell deforms in 1D.
This assumption holds exactly when cells confined to narrow channels elongate
while maintaining constant thickness. In more general settings, the
approximation is not exact.
%
%We consider the case that both membrane and cytosol size is similarly affected by GTPase-caused protrusion or retraction (e.g.\ in concert with aquaporin-facilitated influx or outflux of water \citep{Zehnder2015}).
The opposite extreme case, that cell volume is constant (simply redistributed in
another dimension) whereas membrane size expands/contracts is discussed in
\cref{sec:const_vol}.

We first adopt notation to track material points in the cell. Let $\Gamma_0$
denote the reference domain at time $t=0$. Coordinates in the deformed domain
are in lowercase, i.e., $x$, and coordinates in the original domain are in
capitals i.e.\ $X$. For convenience, we express location and size in terms of
the cell's centroid and radius:
\[
  x_c(t) = \frac{x_1(t) + x_2(t)}{2}, \quad R(t) = \frac{x_2(t) - x_1(t)}{2}.
\]

The mapping between a material point $x$ in the deformed domain, $\Gamma(t)$,
and the original site $X$ in the cell at time $t=0$ is
\[
 x = A(X, t) = x_c(t) + (X - X_c) \frac{R(t)}{R(0)}.
\]
This simplification assumes that the deformation is uniform throughout
the cell (i.e.\ each small volume element of the domain grows/shrinks at the
same rate). The speed of the material point $x$ is
\[
\begin{split}
 \mathbf{v}(x,t) &= \frac{\partial x}{\partial t} =
   x_c'(t) + (X - X_c) \frac{R'(t)}{R(0)} \\
   &= x_c'(t) + (x - x_c(t)) \frac{R'(t)}{R(t)}.
\end{split}
\]
The velocity of flow of material points is thus decomposed into motion of the
cell's centroid  ($x_c'(t)$) and expansion or contraction of its radius
$R'(t)$. (The rate of change of cell length is $L'(t)=2R'(t)$.)

A generalized mass balance equation is obtained following the development by
\citep{Crampin1999}, as shown in our Appendix.  The velocity of material points
enters into the continuity equation in the deforming domain. After some standard
derivation shown (for pedagogical reasons) in the Appendix, we obtain the
following reaction-transport-diffusion equation
\begin{equation}\label{Eqn:RD_Var_Domain}
 u_t + \nabla \cdot \lb \mathbf{v} u \rb = D \Delta u + f(u),\ \mbox{in}\ [x_1(t), x_2(t)],
\end{equation}
with boundary conditions on $\partial\Gamma(t)$ given by
\[
  -D \nabla u + \mathbf{v} u = 0,\ \mbox{on}\ \partial\Gamma(t).
\]
There are two new terms that appear, namely an effective ``advection term'':
$\mathbf{v} \cdot \nabla u$ which corresponds to elemental volumes moving with
the flow due to local growth and a dilution term, $u \nabla\cdot \mathbf{v}$ due
to the local volume change.

For example, if only the centroid moves e.g.\ $x_c(t) = \beta t$, and $R(t) =
R_0$. Then $\mathbf{v} = \beta$, so no dilution takes place. Alternately, if
only the radius of the cell increases with $x_c(t) = 0$,
and $R(t) = R_0 \exp\lb k t\rb$, then $\mathbf{v}(x,t) = k x$.
In this case we have both advection and diffusion. In particular,
\[
   \mathbf{v} \cdot \nabla u = k x \nabla u,
\]
and
\[
   u \nabla\cdot\mathbf{v} = u k.
\]
Refer to \cref{Fig:SingleGTPaseStaticAndGrowingDomain} for an example of this
case.

%%%%%%%%%%%%%%%%%%5
\subsection{Mapping to a constant domain for numerical simulations}

To create numerical simulations of our 1D model (while avoiding the complexities
of a moving boundary problem), it proves helpful to map the moving boundary
problem~\eqref{Eqn:RD_Var_Domain} defined on $\Gamma(t)$ to a stationary
domain $\bar{x} \in [-1, 1]$.  This can be achieved by the change of variables
\begin{equation}\label{Eqn:ChangeOfVars}
 (x,t) \mapsto (\bar x, \bar t) = \left(\frac{x - x_c(t)}{R(t)}, t \right).
\end{equation}
In \cref{sec:numerical_details}, we show that the resulting equation for $u$ is
\begin{equation}\label{Eqn:MainEquation}
    u_{\bar t} = \frac{D}{R^2(t)}u_{\bar{x}\bar{x}} + f(u) - u \lb\frac{\dot
    R(t)}{R(t)}\rb,\ \mbox{in}\ [-1,1].
\end{equation}
Interestingly, the change of coordinates eliminates the advection term, while keeping the
dilution term. At the same time, the diffusion coefficient becomes time
dependent. The advantage of equation~\eqref{Eqn:MainEquation} is that it is
formulated on a fixed domain. This makes its numerical solutions using a method
of lines approach straightforward compared to equation~\eqref{Eqn:RD_Var_Domain}
(see \cref{sec:numerical_details}).

The single GTPase model, so transformed to the domain $[-1,1]$, then becomes
\begin{subequations}
\label{eq:GTPaseOn[-1,1]}
 \begin{alignat}{2}
  \mbox{Active:}&\qquad G_{\bar t} &&= \frac{D}{R^2(t)}G_{\bar{x}\bar{x}} + A(G)G_i - IG - \mathscr{D}(t) G ,\ \mbox{in}\ [-1,1],\\
  \mbox{Inactive:}&\qquad (G_i)_{\bar t} &&= \frac{D_i}{R^2(t)}({G_i})_{\bar{x}\bar{x}} -A(G)G_i +IG - \mathscr{D}(t)  G_i, \ \mbox{in}\ [-1,1],
\end{alignat}
\end{subequations}
where
\[
  \mathscr{D}(t)=\lb \frac{{\dot R}(t)}{R(t)}\rb=\lb \frac{{\dot L}(t)}{L(t)}\rb
\]
is an effective ``dilution term'', and where
$A(G)$ is the activation rate, as before.  In the case that cell volume is
simply redistributed, while only the membrane component expands or contracts,
the term containing $\mathscr{D}(t)$ would be omitted from the second equation (see
also \cref{sec:const_vol}). The dilution term can be expressed alternately in
the above radius or length-dependent forms, used
interchangeably in what follows.

%%%%%%%%%%%%%%%%%%%%
\subsection{Special case: spatially uniform GTPase}
\label{subsec:Uniform_results}

In the case that GTPase activity is spatially uniform inside a cell, we can
reduce model~\eqref{eq:GTPaseOn[-1,1]} to a set of ODEs
\[
\begin{split}
    \frac{\dd G}{\dd t}
        &= \lb \beta + \gamma \frac{G^n}{1 +
        G^n} \rb G_i - IG - G \lb\frac{\dot{L}(t)}{L(t)}\rb,\\
    \frac{\dd G_i}{\dd t}
        &= IG - \lb \beta + \gamma \frac{G^n}{1 + G^n} \rb G_i
        - G_i \lb\frac{\dot{L}(t)}{L(t)}\rb.\\
\end{split}
\]
Using mass conservation we reduce this set of two equations to one equation
by eliminating the inactive GTPase variable:
\begin{equation}
  G_i = \frac{G_T}{L(t)} - G, \label{eqn:mass_conserv}
\end{equation}
where $G_T$ is the total GTPase. This leads to
\begin{equation}
\label{Eq:GTPase_Reduced_to_Well_Mixed}
   \frac{\dd G}{\dd t} = \lb \beta + \gamma \frac{G^n}{1 + G^n} \rb
    \lb \frac{G_T}{L(t)} - G \rb -I G - G \lb\frac{\dot{L}(t)}{L(t)}\rb.
\end{equation}
We later show that this case has interesting consequences when coupled with mechanics. A model of this type, but without the dilution effect was also studied in \citep{holmes2016}.

\section{Results (1): Chemistry only (no mechanics)}
\label{sec:Results1_chem}

We first consider results when domain size affects the signaling, but without
the effect of signaling on the motion of the cell edges. We show how the classic
wave-pining is modified in such cases.

\subsection{Parameter values}

Since some of our results require numerical simulations, we need biological
parameters for the GTPase signaling equations. Typical parameter for the GTPase
dynamics are shown in \cref{Table:Parameters}.

\begin{table}[hbp]\centering
\begin{tabular}{llll}
    \toprule
    \textbf{Parameter} & \textbf{Symbol} & \textbf{Value Range} & \textbf{Reference} \\
    \midrule
%    Active GTPase concentration     & $G$ [$\mu M$]   & $1$ & \citep{Michaelson2001,Mori2008,Jilkine2011} \\
%    Inactive GTPase concentration   & $G_i$ [$\mu M$] & $1$ & \citep{Michaelson2001,Mori2008,Jilkine2011} \\
    Basal activation rate & $\beta$ [\si{\per\second}] & $0.02-0.1$ &
    \citeauthor{Jilkine2011,holmes2016} \\
    Auto-activation rate       & $\gamma$ [\si{\per\second}] & $1-25$ &
    \citeauthor{Jilkine2011,holmes2016} \\
    Decay-rate            & $I$ [\si{\per\second}] & 1 & \citeauthor{Jilkine2011} \\
    EC$_{50}$ parameter   & $G_0$   [\si{\micro\Molar}] & $1$ & \citeauthor{Michaelson2001,Mori2008,Jilkine2011} \\
    Active GTPase diffusion & $D$ [\si{\micro\meter\squared\per\second}] & $0.1$ & \citeauthor{Jilkine2011} \\
    Inactive GTPase diffusion & $D_i$ [\si{\micro\meter\squared\per\second}] & $100$ & \citeauthor{Jilkine2011} \\
    Total GTPase & $G_T$ [\si{\micro\Molar}] & $1-100$ & \\
    Hill coefficient & $n$ & $3-25$ & \citeauthor{holmes2016}\\
    \bottomrule
\end{tabular}
    \caption{Parameters for the GTPase intracellular signaling models.
    %Note that nanomolar is equivalent to $10^{-9}$ moles /L, and $1nM = 10^{-12}$ millimoles /    $\mu L$.
    Note that $G_0$ is the  GTPase level for half-max GTPase autoactivation in the
    Hill function of equation~\eqref{Eq:GTPase_Reduced_to_Well_Mixed}, prior to
    non-dimensionalization. A range of up to 1000-fold has been assumed for the
    diffusion of GTPase protein in the membrane (active) and cytosol (inactive).
    The upper bound for the Hill coefficient is chosen such that there is a
    sufficiently large regime in which oscillations are possible.
    }\label{Table:Parameters}
\end{table}

%%%%%%%%%%FIGURE%%%%%%%%%%%
\begin{figure}[h]\centering
    \includegraphics[width=0.95\textwidth]{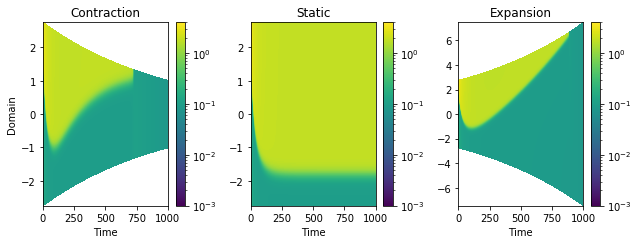}
    \caption{Effect of domain size on polarization. The GTPase model with
    changing length (simulated using the mapping to $[-1,1]$ given by
    \eqref{eq:GTPaseOn[-1,1]}). Initial length is $L = 55 \si{\micro\metre}$ and
    parameters values: $n=4, D_i = 1, D = 0.01, \gamma=2, b=0.1, I = 1, G_T =
    2.5$. Left: A shrinking domain with $R(t) = R_0 \exp\lb-k t \rb$, as the
    cell shrinks $G_T$ increases. Middle: A static domain with $R(t) = R_0$.
    Right: A growing domain with $R(t) = R_0 \exp\lb k t\rb$, (with $k=10^{-3}$):
    as the cell grows, $G_T$ decreases. In both cases in which the domain size
    changes, $G_T$ eventually leaves the parameter regime in which wave pinning
    is possible. The transition is very rapid.}
    \label{Fig:SingleGTPaseStaticAndGrowingDomain}
\end{figure}

%
% NOT RELEVANT FOR THIS PAPER!
%
% The concentration of external gradients and the internal concentrations of Rac /
% Rho  differ by 2-3 orders of magnitude. Following \citep{Mori2008,Jilkine2011} we
% can roughly estimate the internal concentrations of GTPases to be 1 $\mu$M.
% External concentrations are of the order of nM (nanomolar).
%
% Cells can detect a very shallow gradient, down to 2\% across their diameter.
% Their internal gradient is greatly amplified, producing a much larger
% macroscopic gradient inside the cell. See \citep{Jilkine2011} and references
% therein.
%
%%%%%%%%%%%%%%%%%%%%%%%%%%%%%%%%%%
%%%%%%%%%%%%%%%%%%%%%%%%%%%%%%%%%%
\subsection{GTPase dynamics in constant and fluctuating cell size}

In previous work \citep{holmes2016}, predictions of
Eqs.~\eqref{Eqn:General_GTP_model} was characterized in a static 1D domain using
several approximation methods, including Local Perturbation Analysis (LPA).
Coexistence of several types of steady state solutions were found.  Here, we
extend the results to domains with fluctuating size.

We first simulated the GTPase PDE model \eqref{eq:GTPaseOn[-1,1]} on a domain
whose length is a specified (exponential) function of time. Initial parameter
values allow for wave-pinning in each case. Results are shown in panels of
\cref{Fig:SingleGTPaseStaticAndGrowingDomain} with yellow tint for the high
GTPase activity.

In both shrinking and growing model cells, the polarization collapses around $T
= 800$. Note the rapidity of the transition from polarized to uniform activity.
Loss of polarization if the cell becomes too small is consistent with
bifurcation results of \citep{mori2011}. Loss of polarization when the cell grows
too much corresponds to the case where mean total GTPase concentration in the
cell drops, also consistent with \citep{mori2011}. In short, either increase or
decrease of cell size eventually moves the system out of the regime of
parameters in which wave-pinning can be sustained.

Up to this point, we considered only the GTPase distributions inside the cell,
and the effect of changes in cell size on that internal level. To describe the
feedback between chemistry and mechanics, we next endow the cell with additional
properties that link its size to mechanical forces. The basic ideas follow those
of \citep{Zmurchok2018}, but with the adjusted mass balance in place.

%%%%%%%%%%%%%%%%%%%%%%%%%%%%%%%%%%%%%%
%%%%%%%%%%%%%%%%%%%%%%%%%%%%%%%%%%%%%%
\section{Model adaptation (2): Cell mechanics}
\label{sec:ModelAdapt_mechanics}

We now specify assumptions and equations for the endpoints of the cell, which
then determine its size.  As shown in \cref{fig:modelGeometry}, the cell is
modeled as a 1D contractile element of length $L$ and rest length $L_0$,
analogous to a Kelvin-Voigt element with constitutive relation for the stress,
$\sigma$ given by
\[
    \sigma(t) = E \epsilon(t) + \eta \frac{d \epsilon}{dt},
\]
where $E$ is the cell's Young modulus,  $\eta$ the viscous component and
$\epsilon$ the strain.  Cell motion is in an overdamped regime where inertial
terms in the Newtonian force balance are negligibly small. In this regime, the
equation of motion becomes $F = \mu v$, where $\mu$ is the cell's frictional
coefficient. We denote the forces applied at nodes $x_1, x_2$ by $F_1, F_2$
respectively. The cell's strain is given by
\[
    \varepsilon = L - L_0=x_2 - x_1 - L_0.
\]

\begin{table}[hbtp]\centering
\begin{tabular}{llll}
    \toprule
    \textbf{Parameter} & \textbf{Symbol} & \textbf{Value Range} & \textbf{Reference} \\
    \midrule
    End point friction & $\mu$  [\si{\pico\newton\second\per\micro\metre}] &
    $10^{3} - 10^{5}$ & \citeauthor{Drasdo2005} \\
    Cell viscosity & $\eta$  [\si{\pico\newton\second\per\micro\metre}]  & $6
    \times 10^{1}$ & \citeauthor{Palsson2008} \\
    Cell Young's modulus & $E$ [\si{\pico\newton\per\micro\metre}] & $5 \times
    10^{3}$ & \citeauthor{Drasdo2005,Palsson2008} \\
    Cell rest length & $L_0$  [\si{\micro\metre}] & $10-60$& \\
    Maximum cell force     & $F_{\text{max}}$ [\si{\pico\newton}] & $10^{3} - 10^{5}$ & \\
    Cell force half-rate constant        & $G_c$ [\si{\micro\Molar}] & $1.25 G_0$ &  \\
    Cell force Hill coefficient          & $m$ & $4-10$ &  \\
    \bottomrule
\end{tabular}
\caption{Mechanical parameters for the model cell. $G_0$ is defined in \cref{Table:Parameters}. Parameter values from 3D models are used to  inform our 1D model.
}\label{Table:MechPars}
\end{table}

%%%%%%%%%%%%%%%%%%%%%%%%%%%%%%
\subsection{Equations for the cell edges}

Given the assumed cell mechanics, the front and rear edge coordinates, $x_1, x_2$, satisfy the following equations of motion
\begin{subequations}
\label{Eqn:CellKV}
\begin{align}
   \mu \frac{\dd x_1}{\dd t} &= F_1 + E(x_2 - x_1 - L_0) + \eta\lb \frac{\dd x_2}{\dd t} - \frac{\dd x_1}{\dd t} \rb, \\
   \mu \frac{\dd x_2}{\dd t} &= F_2 - E(x_2 - x_1 - L_0) - \eta\lb \frac{\dd x_2}{\dd t} - \frac{\dd x_1}{\dd t} \rb.
\end{align}
\end{subequations}

Solving the above equation for the time derivatives, we obtain two differential
equations for the cell's endpoints
% \[
%     \begin{pmatrix} \dot{x}_1 \\ \dot{x}_2 \end{pmatrix}
%            = \frac{1}{\mu(\mu + 2 \eta)}
%            \begin{pmatrix} \mu + \eta & \eta \\ \eta & \mu + \eta \end{pmatrix}
%            \begin{pmatrix} F_1 + E \varepsilon \\ F_2 - E \varepsilon \end{pmatrix}.
% \]
% Thus we get
\begin{equation}\label{Eqn:MechanicalFinal}
    \begin{pmatrix} \dot{x}_1 \\ \dot{x}_2 \end{pmatrix} = \frac{1}{\mu(\mu + 2 \eta)}
           \begin{pmatrix} \mu(F_1 + E\varepsilon) + \eta (F_1 + F_2) \\
                           \mu(F_2 - E\varepsilon) + \eta (F_1 + F_2)
           \end{pmatrix}.
\end{equation}

%%%%%%%%%%%%%%%%%%%%%%%%%%%%
\subsection{Equivalent system for centroid and radius}
We can also write this system in terms of the cell's centroid,
$x_c(t)=(x_1 + x_2) / 2$, and cell's radius, $R(t)=L/2=(x_1 - x_2) / 2$,
\[
    \begin{pmatrix} \dot{x}_c \\ \dot{R} \end{pmatrix} = \frac{1}{2\mu(\mu + 2 \eta)}
           \begin{pmatrix} (\mu+2\eta)(F_1 + F_2) \\
                           \mu(F_2-F_1 -2 E\varepsilon)
           \end{pmatrix},
\]
where $ \varepsilon = 2R-L_0$. In simplest form, the latter is written
\[
  \dot{x}_c= \frac{F_1 + F_2}{2\mu} , \quad
  \dot{R} = \frac{F_2-F_1 -2 E (2R-L_0)}{2(\mu+2\eta)}.
\]

From the second equation we recognize the familiar spring-length equation  (also
directly obtained by subtracting \eqref{Eqn:CellKV}(b-a)):
\begin{equation}
  (\mu+2\eta) \frac{\dd L}{\dd t} = (F_2-F_1) -2 E(L - L_0).
\label{eq:dLi/dt}
\end{equation}

In the Appendix, we also show that imposing external forces is equivalent to
changing the rest-length of the spring, as in \citep{Zmurchok2018,Liu2019}.

%%%%%%%%%%%%%%%%%%%%%%%%%%%%
\subsection{Special cases}
From the equations for $x_c(t),R(t)$, it is easy to see that in the case of
equal forces on the two cell ends,  we get
\[
  F_1=F_2=F \quad \Rightarrow \quad \dot{x}_c=F/\mu, \quad \dot{R}=0,
\]
so the cell moves at speed $F/\mu$ without deforming.  In the case of equal and
opposite forces on the two cell ends
\[
  F_1=-F_2=|F| \quad \Rightarrow \quad \dot{x}_c=0, \quad \dot{R}=(F - E\varepsilon)/(\mu+2\eta).
\]
Hence, the cell does not move, but its size changes.

%%%%%%%%%%%%%%%%%%%%%%%%%%%%
\section{Model adaptation (3): Coupling mechanics and chemical signaling}
\label{sec:ModelAdapt3}

We now link signaling inside the cell to cell mechanical properties. In
\citep{Zmurchok2018}, this coupling was done for the spatially uniform GTPase case without the dilution terms. We
extend that model by considering a fully nonuniform GTPase distribution inside a cell correct mass balance for cell size changes.

%%%%%%%%%%%%%%%%%%%%%%%%%%%%
\subsection{GTPase activity affects cell contraction and expansion}

The Rac-induced cell protrusion and Rho-induced cell contraction can
be modeled in various ways.  Here we will assume that Rac and Rho (by assembling
actin or activating actomyosin) create forces at cell endpoints. These forces
either push the endpoint outwards (Rac) or pull it inwards (Rho). Forces are taken
to be a function of the local GTPase level. In the spatially dependent GTPase
cases, each cell endpoint can experience a distinct force.

Denote by $G \in \{ G_R, G_\rho \}$ the active form of either GTPase. Then the
forces $F_{1,2}$ on the cell's endpoints are taken to be
\begin{equation}
 F_k(t) = \pm F_{\text{max}} \frac{G^m(x_k, t)}{G_c^m + G^m(x_k, t)},
    \label{eq:HillFnForce}
\end{equation}
where $k=1, 2$. The sign of the force would correspond to an outwards force for
Rac and an inwards force for Rho.

According to \eqref{eq:HillFnForce}, force builds up continuously as $G(x_k)$
approaches a characteristic level, $G_c$.  The force saturates to a
maximal level $F_{\text{max}}$ and the sharpness of the response increases with
$m$. (Note that our framework here appears to be different from
\citep{Zmurchok2018,Liu2019}, where GTPase activity affected the cell rest-length;
we show the mathematical equivalence of these two model variants in the Appendix.)

%%%%%%%%%%%%%%%%%%%%%%
\subsection{Complete model equations}

In summary, the full model consists of two PDEs for the active and
inactive GTPase distributions (with Neumann boundary conditions), equations for
the cell centroid and radius, and a GTPase-dependent set of forces on the
(transformed) cell ends. That is, the system is as follows:
\begin{subequations}\label{eq:FullModelsEqs}
\begin{align}
 &\mbox{Active GTPase:}\quad   G_{\bar t} = \frac{D}{R^2(t)}G_{\bar{x}\bar{x}} + f(G, G_i) - G \lb\frac{\dot
    R(t)}{R(t)}\rb,\ \mbox{in}\ [-1,1],\label{eq:active_gtpase} \\
 & \mbox{Inactive GTPase:}\quad   {(G_i)}_{\bar t} =
 \frac{D_i}{R^2(t)}{G_i}_{\bar{x}\bar{x}} - f(G,G_i) - G_i \lb\frac{\dot
    R(t)}{R(t)}\rb,\ \mbox{in}\ [-1,1],\label{eq:inactive_gtpase}\\
&\mbox{Kinetics, BCs:}\quad      f(G, G_i)  = \left( \beta + \gamma \frac{G^n}{1 + G^n}\right)G_i-IG,
    \quad G_{\bar{x}}= {G_i}_{\bar{x}}=0\ \mbox{for}\ \bar{x}=\pm1,  \label{eq:kinetics}
      \\
&\mbox{Cell centroid, radius:}\quad      \dot{x}_c= \frac{F_1 + F_2}{2\mu},\qquad
    \dot{R} = \frac{F_2-F_1 -2 E (2R-L_0)}{2(\mu+2\eta)},\label{eq:cell_length}\\
&\mbox{Forces on cell ends:}\quad       F_1 = \pm F_{\text{max}} \frac{G^m(-1)}{G_c^m + G^m(-1)},
  \quad F_2 = \pm F_{\text{max}} \frac{G^m(1)}{G_c^m + G^m(1)}.\label{eq:cell_forces}
\end{align}
\end{subequations}

In the following sections, we study variants of this model with and without the
diffusion terms, and with the  protrusive (Rac, $G_R$) and contractile (Rho,
$G_\rho$)  GTPases acting on their own and together.

%%%%%%%%%%%%%%%%%%%%%5
\section{Results (2): Spatially uniform GTPase linked to forces and dynamic cell size}
\label{sec:uniform_results}

In this case, we drop the diffusion terms in the first two PDEs in
Eqs.~\eqref{eq:FullModelsEqs}, so that active GTPase $G$ is time dependent only,
uniform throughout the cell.  The conservation statement for GTPase reduces to
\[
  \int_0^{L(t)} (G(t)+G_i(t))dx = L(t) [G(t)+G_i(t)]= G_T >0
\]
where $G_T$ is the constant total amount of GTPase in the cell. Since $G, G_i>0$ we can state that
\[
  G(t) < \frac{G_T}{L(t)}.
\]
This implies that the region of interest in the $L-G$ plane is bounded
above by the curve $G=G_T/L$. Indeed we show below
(\cref{Lemma:InvariantRegion}) that this forms an invariant region, i.e.\ that
trajectories do not cross this curve.

Furthermore, since $G$ is spatially uniform, and hence the same at both cell
ends, we have that forces on the ends are equal in magnitude but opposite in sign
i.e.\ $F_2 = -F_1$, and the cell tension, $F_T(G) = F_2(G) - F_1(G)$, satisfies
$|F_T| = 2|F_1|$. This also means that the cell's centroid remains fixed i.e.\
$\dot{x_c} = 0$.

Coupling the length of the cell to the GTPase dynamics,
Eqn.~\eqref{Eq:GTPase_Reduced_to_Well_Mixed}, leads to the reduced model variant
to study:

\begin{subequations}\label{Eq:DynSysUnif}
\begin{align}
    \frac{\dd L}{\dd t} &= \frac{-2E(L - L_0) + F_T(G)}{\mu + 2 \eta}, \label{Eq:dL/dt_uniform_G}\\
    F_T(G) &= \pm 2F_{\text{max}} \frac{G^m}{G_c^m + G^m},\label{eq:Force_uniform_G}\\
    \frac{\dd G}{\dd t} &= \lb b + \gamma \frac{G^n}{1+G^n}\rb\lb \frac{G_T}{L} - G\rb - IG -
       G\lb \frac{\dot{L}}{L}\rb. \label{Eq:dG/dt_uniform_G}
\end{align}\label{sys:uniform_G}
\end{subequations}
The sign on the force terms in \eqref{eq:Force_uniform_G} is $+$ for a
protrusive GTPase ($G=G_R$) and $-$ in the contractile GTPase case ($G=G\rho$).
We have reduced the original system \eqref{eq:FullModelsEqs} to the two coupled
ODES \eqref{Eq:dL/dt_uniform_G} and \eqref{eq:Force_uniform_G} for $L(t), G(t)$.
This system can be studied in the $L-G$ phase plane. Below we explore the
detailed behavior of this system.

To ensure that the cell length remains positive in the contractile (Rho) case we
require that $E L_0 / F_{\text{max}} > 1$. In addition, we assume that in this
section the force switches ``ON'' at  $G_c > 1$; the case in which $G_c < 1$
follows readily.

\begin{result}[Invariant Region]\label{Lemma:InvariantRegion}
    The region defined by
    \[
        \Delta \coloneqq \{ (L, G) : L > 0;\ G \geq 0;\ G \leq G_T / L \}
    \]
    is invariant with respect to the flow given by equations~\eqref{Eq:DynSysUnif}.
\end{result}

\begin{proof}
Since $E L_0 / F_{\text{max}} > 1$, we have that $L' > 0$ for $L >0$, and
similarly $G' > 0$ whenever $G=0$. So flow cannot leave the region through the
axes. Now consider the direction of the  flow along the boundary curve $f(L) =
G_T /L$. The normal vector along the curve is given by $(1, L^2 / G_T)$, then
computing the component of flow along this normal direction, we find
\[
    \lb L'(t), G'(t) \rb \cdot (1, L^2 / G_T) < 0.
\]
so the flow is {\it into}\ the region, and cannot leave through this boundary.
Hence flow is trapped in the region $\Delta$, which is thus invariant.
\end{proof}

\begin{figure}\centering
    \input{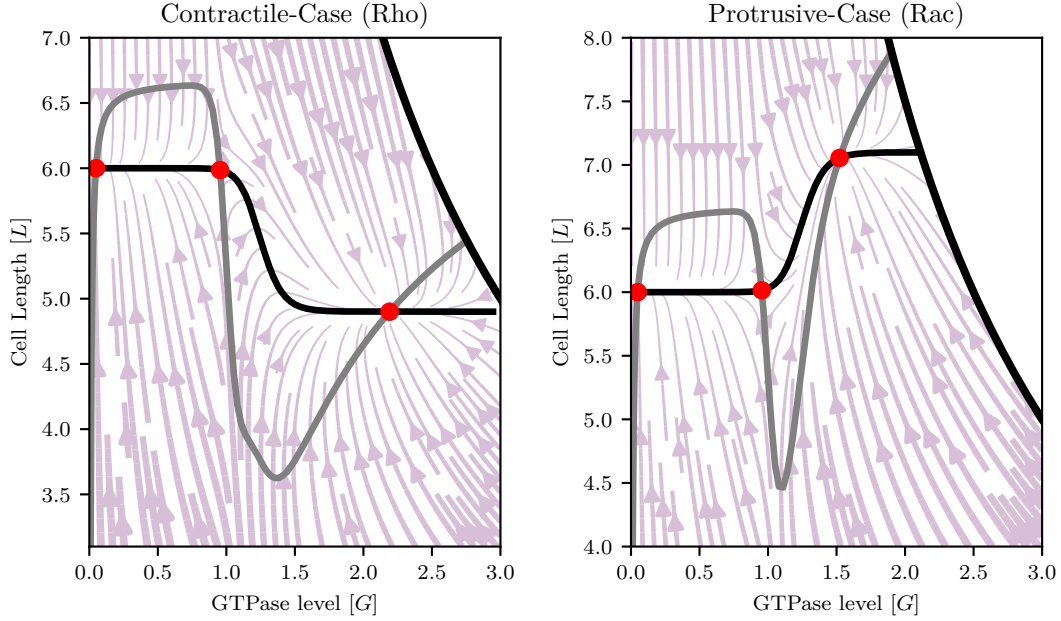}
    \caption{Phase-plane plots for the contractile (left: Rho) and protrusive
    (right: Rac) cases of the dynamical system~\eqref{Eq:DynSysUnif}. In solid
    black $L$ nullcline, and in grey the $G$ nullcline. The steady states are shown
    by red dots. Note the invariant region, bounded by the curve $L=G_T/G$. Here
    $G_T = 15, G_c = 1.25, \gamma = 2.5, n = 25, m = 16, L_0 = 6$, and other
    parameter values are as in Table~\ref{Table:Parameters}, and
    \cref{Table:MechPars}. In both cases, the most left ($G_1$) and most
    right ($G_3$) steady states can be approximated using a sharp switch
    approximation, the middle steady states are saddles.
    }\label{Fig:PhasePlane}
\end{figure}

In \cref{Fig:PhasePlane}, we show typical phase plane behavior of the
system~\eqref{Eq:DynSysUnif} for the contractile Rho ($G=G_\rho$, left) and
protrusive Rac ($G=G_{R}$, right) cases. In both cases we find the region
$\Delta$ bounded by the (thick black) curve $G= G_T / L$. The level of total
GTPase, $G_T$ controls the position of this curve, so that the region is larger
for larger values of $G_T$.  The nullclines, obtained by setting $\dd L/\dd t=0$
in \eqref{Eq:dL/dt_uniform_G} and $\dd G/\dd t=0, {\dot L}=0$ in
\eqref{Eq:dG/dt_uniform_G} have shapes with the following features: the $L$
nullcline (black) is sigmoidal, but its orientation flips between the two cases.
The $G$ nullcline  (grey) is ``Z-shaped''. It remains to determine how many
possible steady states can occur at intersections of such curves, and which
parameters govern possible bifurcations.

%%%%%%%%%%%%%%%%%%%%%%%%%%
\subsection{Sharp-switch analysis}

To help explore this question, we use a common shortcut, the sharp-switch
approximation \citep{holmes2016} in the $L$ equation. Next we estimate the number
of possible steady states.
%
% In the next result we show
% that there is an upper bound for the number of steady states.

\begin{result}[Number of steady states]\label{Lemma:NumberOfStSt}
    Consider ODE~\eqref{Eq:dG/dt_uniform_G} with $L(t)$ set to either of the two steady
    states of the ODE~\eqref{Eq:dL/dt_uniform_G}. Then ODE~\eqref{Eq:dG/dt_uniform_G}
    has either one or three positive steady states.
\end{result}

\begin{proof}
    In the contractile (Rho) case, the  $L$ nullcline is given by:
    \[
        L_{eq} = L_0 - \frac{F_{\text{max}}}{E} \frac{G^m}{G_c^m + G^m}.
    \]
    In the protrusive (Rac) case, the minus sign is replaced by +.

    We assume that the switch is sharp and approximate $L$ by
    $L_{\infty} \in \{L_0, L_0 - F_\text{max}/E\}$. The nullcline for $G$ can be written as:
    \[
    L[A(G)+I] +{\dot L}= A(G) \frac{G_T}{G}, \quad L, G >0,
    \]
    where $A=(b+\gamma h(G))$ and $h(G)$ is the Hill-function
    in~\eqref{Eq:dL/dt_uniform_G}.  Substituting the expression for ${\dot L}$
    from the ODE \eqref{Eq:dG/dt_uniform_G}, leads to a more complicated
    nullcline equation which we do not display, but which leads to the sigmoidal
    curve plotted in \cref{Fig:PhasePlane} (black) and in
    \cref{fig:RhoAndRac_nullclines}
    (blue). Observe that the direction of this switch flips between the Rac and
    the Rho cases, as shown in the top and bottom panels of
    \cref{fig:RhoAndRac_nullclines}, respectively.

\begin{figure}\centering
    \input{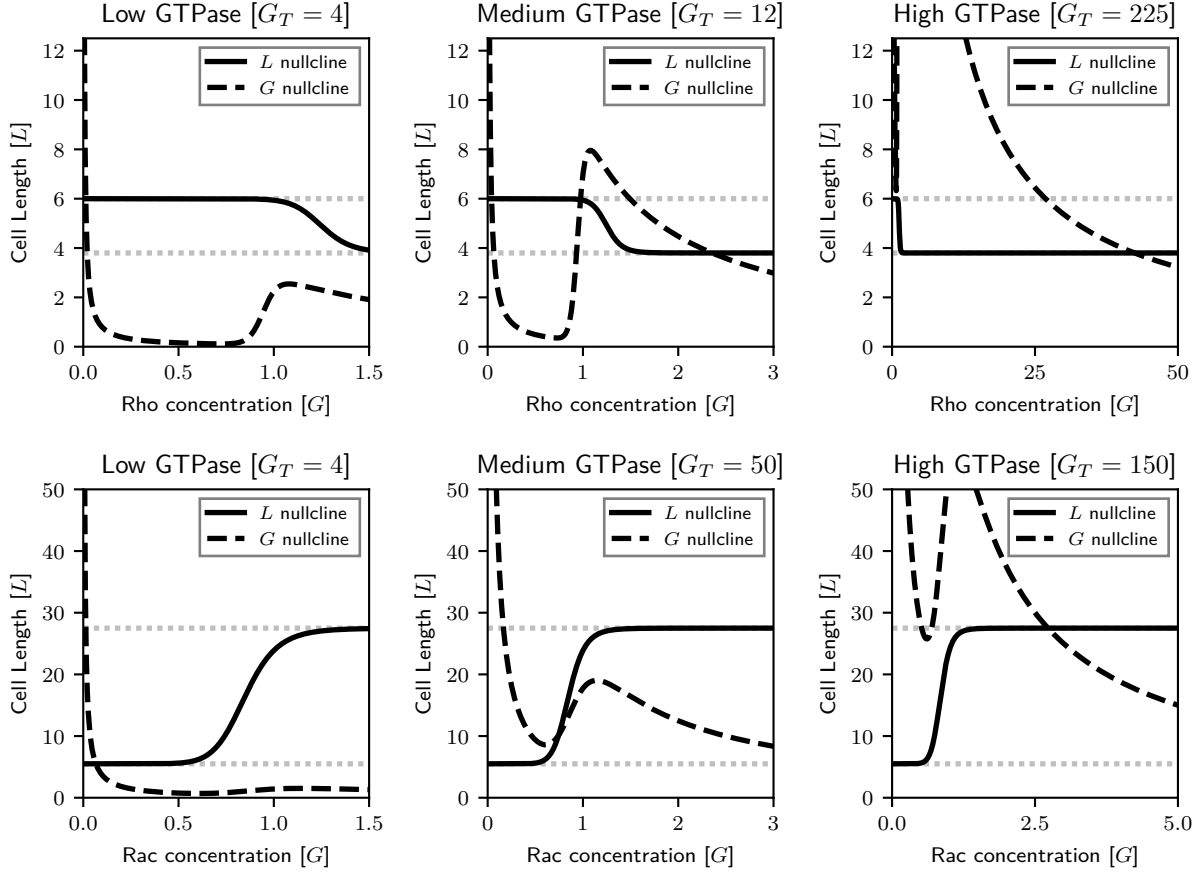}
    \caption{Possible nullcline configurations.
    Top: contractile GTPase (Rho) case;
    Bottom: protrusive GTPase (Rac) case.
    Top: For Rho, the left panel shows the existence of $G_1$.
    The middle panel shows the bistable case,
    i.e.\ two steady states in which the cell is large and GTPase is small
    ($G_1$, and $G_3$). The right panel shows the existence of a
    single steady state $G_3$. Other parameters $\gamma = 2.5,
    b= 0.02, I = 1, L_0 = 6$. The dotted lines are the upper and lower sharp
    switch approximations of the sigmoidal $L$ nullcline.  Bottom: For Rac, the
    middle panel demonstrates the case where a stable limit cycle oscillation is
    expected, a situation that cannot occur in the Rho case.
    Other parameters $\gamma = 0.9, b= 0.1, I = 1, L_0 = 6$. For the values of
    the approximate steady states $G_1, G_2$, and $G_3$ see
    \cref{subsec:SharpSwitch}.
    }\label{fig:RhoAndRac_nullclines}
\end{figure}

    As we are seeking steady states, we may set ${\dot L}=0$, simplifying the resulting equation to

    \[
        L= %\coloneqq
        \frac{G_T\lb b + \gamma h(G)\rb}{G\lb I + b + \gamma h(G) \rb} \equiv n(G),
    \]
   where we have defined $n(G)$ as the rational function that appears.
    We note that asymptotically $n(G)$ behaves as:
    \[
        \lim_{G \to 0} n(G) = \infty, \qquad
        \lim_{G \to \infty} n(G) = 0.
    \]
    To find intersections of $n(G)$ with $L_{\infty}$ we must solve:
    \[
        G^{n+1} L_{\infty}\lb I + b + \gamma\rb - G^n G_T\lb b + \gamma\rb + G L_{\infty} \lb I + b\rb - b G_T = 0.
    \]
    Let $p$ be the number of positive solutions of this polynomial. Since the
    sequence of the polynomial's coefficients features three sign changes, we
    have by Descartes' rule that $3 - p$ must be non-negative and even (for a
    statement of Descartes' rule we refer the reader to Appendix~2 of
    \citep{murray1989}). We conclude that $p = 1$ or $3$.
\end{proof}

\cref{Lemma:NumberOfStSt} implies that in most cases we have either one or three
solutions. Note that in the most general case when $L$ is not set equal to
either of its two steady-states, Descartes' implies that there may be $1,2,5$ or
$7$ steady states. However, since \cref{Lemma:NumberOfStSt} restricts the number
of steady states when $L = L_{\infty}$ these additional steady states must occur
in the thin region $G \sim G_c$ in which $L$ transitions between the two
possible value for $L_{\infty}$. Thus in the vast majority of cases we only have
$1$ or $3$ steady states. See also \cref{fig:RhoAndRac_nullclines}.

The sharp-switch approximation leads to explicit values for some of the steady
states of the system, their eigenvalues, and hence full characterization of
their stability and conditions for existence. The detailed calculations are
provided in Appendix~\ref{subsec:UniformLinStab}. A brief summary is given
in \cref{Tab:StStSummary}.  This approximation does not allow easy computation
of steady states that occur when nullclines intersect along the upswing portion
of the switch. However, the geometry of such intersections allows us to
determine stability nonetheless, using geometric methods described in
\citep{murray1989} (see also \cref{subsec:UniformLinStab} for linear stability
analysis).

\begin{table}[!ht]
    \centering
    \begin{tabularx}{0.9\textwidth}{p{0.2\textwidth} @{} >{}Y p{0.2\textwidth} @{} p{0.4\textwidth} @{} }
    \toprule
        {\bf GTPase Level} & {\bf $L_{eq}$} & {\bf $G_{eq}$} & {\bf Stability} \\
        \midrule
        $G < \min(1, G_c)$ & $L_0$ & $\dfrac{b G_T}{L_0\lb b + I \rb}$   &stable
        node\\[3ex]
        transition &? &? & Rho: saddle. Rac: unstable, saddle \\[3ex]
        %$G_c < G < 1$ & $L_1$ & $\dfrac{b G_T}{L_{eq}\lb b + I \rb}$ & stable node\\[3ex]
        $1 < G < G_c$ & $L_0$ & $\dfrac{(b + \gamma) G_T}{L_0\lb b + \gamma + I
        \rb}$ & stable node\\[3ex]
        transition &? &? & Rho: saddle. Rac: unstable, saddle \\[3ex]
        $G > \max(G_c, 1)$ &  $L_1$ & $\dfrac{(b + \gamma) G_T}{L_0\lb b + \gamma + I \rb}$ &stable node\\
        \bottomrule
    \end{tabularx}
    \caption{Sharp switch approximated steady state summary.
  %  The ``switchstate'' index $i-j$ refers to the On-Off states in the $L$ Hill function($i$) and in the GTPase activation Hill function ($j$).
    Here
    $L_1=L_0-(F_\text{max}/E$) for Rho and $L_1=L_0+(F_\text{max}/E$) for Rac.
    The transitions in the table can contain one or more steady states. In
    case of a single steady state at the transition, that state is unstable.
    Note that the number of coexisting steady states depends
    on the parameters. See text and Appendix~\ref{subsec:UniformLinStab} for
    details.}\label{Tab:StStSummary}
\end{table}

In the contractile GTPase (Rho) case, we find three steady states, all stable
nodes, with possible coexistence of all three (These are separated by steady
states that occur on the portion of the switch not captured by the
approximation). In the protrusive (Rac) case there are either two coexisting
stable steady states (separated by a steady state not captured by the
approximate) or no coexisting stable steady states at all. When no stable steady
states exist we must have a stable limit cycle, since $\Delta$ is invariant.

\subsection{Simulations}
\begin{figure}\centering
\includegraphics[width=0.9\textwidth]{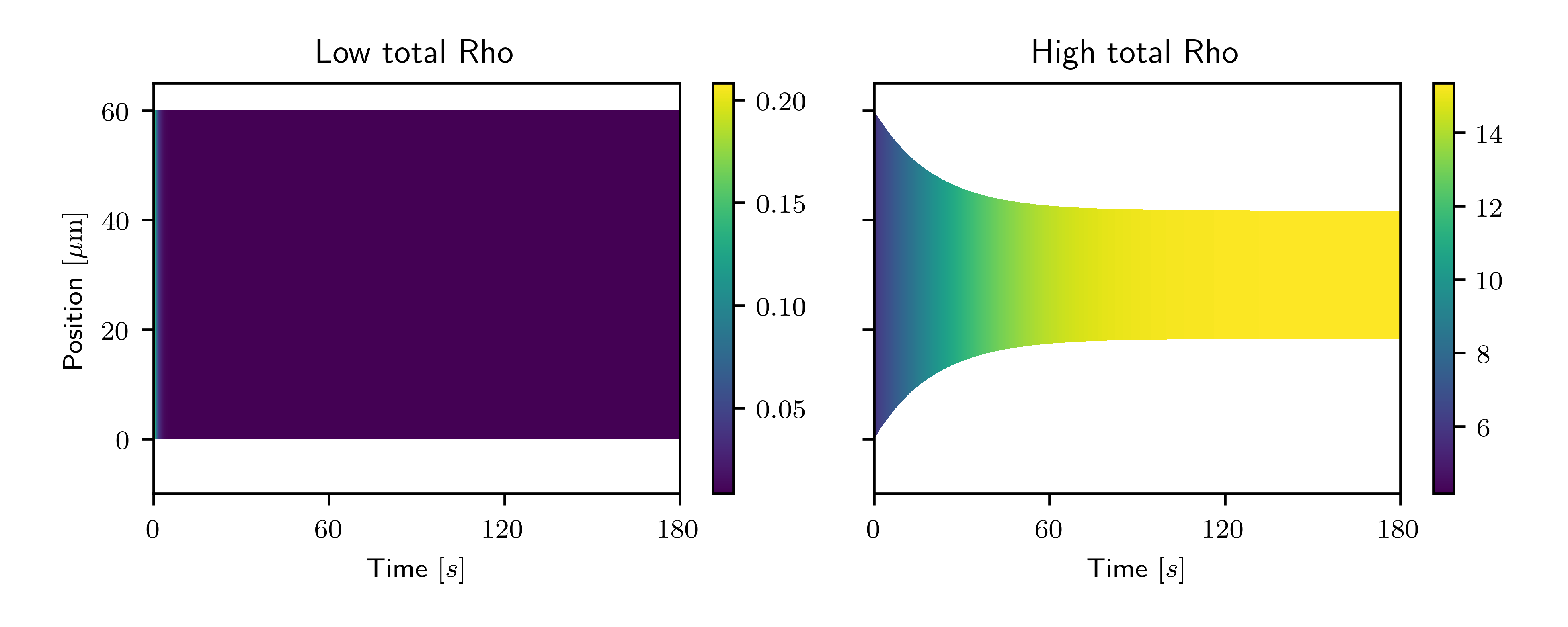}
\includegraphics[width=0.95\textwidth]{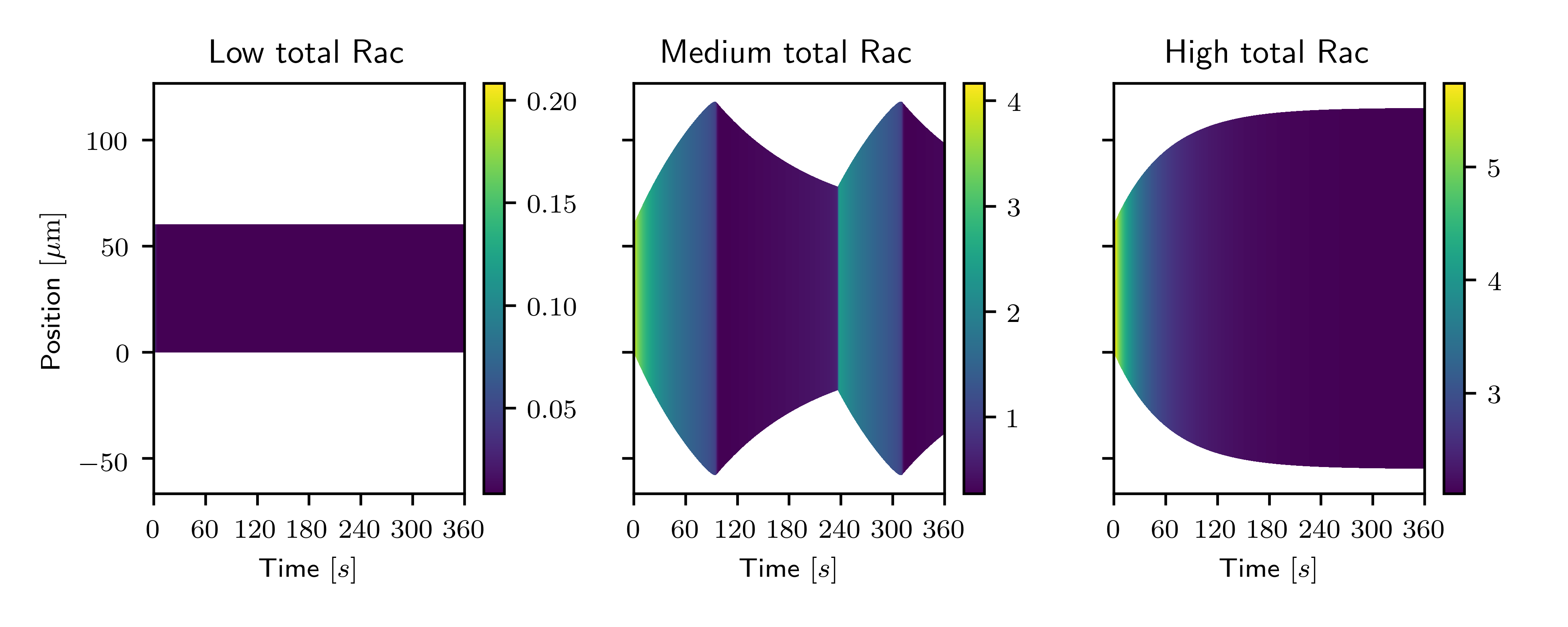}
\caption{Simulation of a cell with uniform internal GTPase.
Top: Rho, the GTPase that leads to contraction.  Bottom: Rac, the GTPase that
leads to cell protrusions. Left panels: Low total amount of GTPase, $G_T = 2.5$:
no change takes place in either case. Right panels: High total GTPase, $G_T
= 50$: the cell becomes uniformly high in Rho/Rac and hence shrinks/expands,
respectively. Bottom Middle: Cell with intermediate total Rac and parameters
such that oscillation is possible. Same parameters as the bottom middle of
\cref{fig:RhoAndRac_nullclines} ($E = 250\
[\si{\pico\newton\per\micro\metre}]$).  Initially, the cell is at its
``rest-length'' $L(0) = L_0$, and GTPase is uniformly set to $G_T / L$. In
the Rac case the cell's stiffness is $E = 500\
[\si{\pico\newton\per\micro\metre}]$, while in the Rho case it is $E = 1500\
[\si{\pico\newton\per\micro\metre}]$. This choice ensures that the cell does not
vanish i.e.\ $E L_0 / F_{\text{max}} > 1$.}\label{fig:RhoRacUniformSim}
\end{figure}

We simulated the model for the spatially uniform Rho, and Rac case (see
\cref{fig:RhoRacUniformSim}), starting with a cell at its rest-length. Results
are shown in the form of a kymograph, with time on the horizontal axis, and a
profile of the cell and its internal active GTPase concentration on the vertical
axis.

In both the Rho and Rac case of the single-GTPase model, when the total amount
of the GTPase, $G_T$ in the domain is too low (left panels), there is no
response.  For high $G_T$ (right panels), the system is triggered to switch to
the high GTPase steady state (shaded yellow). For Rho (top panels in
\cref{fig:RhoRacUniformSim}), the cell contracts as a result, and the high
activity state becomes amplified even more due to the smaller volume in which it
is concentrated. In contrast, in the Rac case (Bottom panels of
\cref{fig:RhoRacUniformSim}), expansion caused by high Rac results in a larger
cell, but also dilutes Rac activity, depressing it to a lower level.

In the Rac, but not the Rho case, the conflicting effects of expansion and
dilution can set up an oscillatory behavior (bottom middle panel,
\cref{fig:RhoRacUniformSim}). The cell alternates between long and short
phases, with fluctuations in its Rac activity.

%%%%%%%%%%%%%%%%%%%%%%%%%%
\subsection{Bifurcation diagrams}

\begin{figure}\centering
    \includegraphics[width=1.1\textwidth]{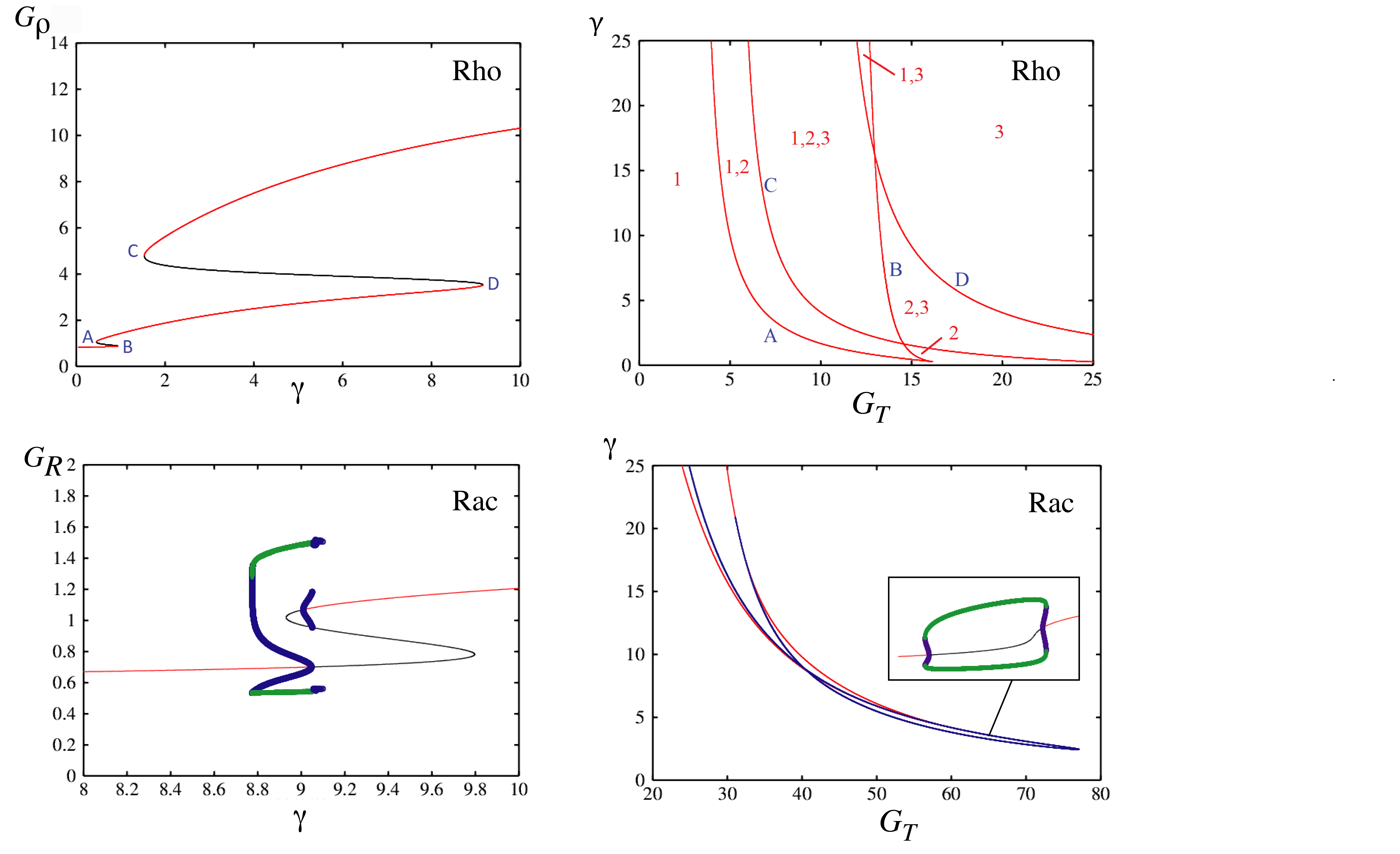}
    \caption{Bifurcation diagrams for the spatially uniform GTPase cases.
    Top: the contractile (Rho) case, showing the possibility of tristability.
    Bottom: The protrusive (Rac) case.
    Left panels: one-parameter bifurcation with respect to the feedback
    activation rate, $\gamma$. Right panels: two-parameter bifurcation plot with
    respect to total GTPase, $G_T$ and $\gamma$.  The fold bifurcations in the
    Rho case (top left) labeled A, B, C, D, are tracked by curves with the same
    labels in the two-parameter plot (top right). Numbers 1,2,3 indicate
    existence of the corresponding steady states in regions of the same plot.
    (For example,  ``1,2,3" indicates existence of all three equilibria $G_1,
    G_2, G_3$, as discussed in \cref{subsec:UniformLinStab}. The parameters are
    $b=1, I=5, m=n=20, L_0=3, \frac{2E}{\mu + 2
    \eta}=5, \frac{2 F_{max}}{\mu+2 \eta}=10, G_c=4, G_T=15$.
    Bottom: Left: a supercritical Hopf bifurcation is evident at $G_T=40$.
    Right: the inset shows a limit cycle that exists in the range of $G_T=65$
    and $3.2<\gamma<3.6$. Limit cycles are possible only in the Rac case. The
    parameters are $b=1, I=10, m=n=8, L_0=5.5, \frac{2E}{\mu + 2 \eta}=5,
    \frac{2 F_{max}}{\mu+2 \eta}=55, G_c=0.85$.}\label{Fig:BothBifurcYLiu}
\end{figure}

Finally, we characterized the regimes of behavior in the spatially uniform cases
by creating bifurcation plots of model~\eqref{Eq:DynSysUnif}. Results are shown
in panels of \cref{Fig:BothBifurcYLiu} for Rho (top) and Rac (bottom). It is
evident that a region of tristablity can exist in the Rho case (labeled C),
whereas the Rac case has a regime in which limit cycle oscillations can exist.
Panels on the left should be compared with the 2-parameter bifurcation plot for
the single-GTPase case (Fig.~3a) in \citep{holmes2016}. The distinction here is
that we have taken cell size into account, which increases the number of
possible steady states and introduces the possibility of Hopf bifurcations.

%%%%%%%%%%%%
\subsection{Extensions to a Rac-Rho model}
So far, we explored only the single GTPase model, for either Rac or Rho acting
alone. We asked how conclusions would change if both Rac and Rho operate
simultaneously (with mutual inhibition) as in
\citep{holmes2016,Zmurchok2018,Zmurchok2019}. We briefly examined the spatially
uniform case using the same sharp switch approximation.  Rac and Rho each have
two applicable thresholds: one for turning on the self-activation, and another
for the cell length, which implies three possible cases for equilibrium levels.
With two GTPases in the model, there would be nine possible equilibria, (which
may not all coexist in the same parameter regime). There are many parameters to
consider. From preliminary exploration, we are able to find multi-stability and
oscillatory regimes consistent with results for the single GTPase model. We did
not find new behaviors unique to the two-GTPase model, but see also
\citep{Zmurchok2019} for further results.

%%%%%%%%%%%%%%%%%%%%%%%%%%%%%%%%%%%
\section{Results (3): Spatially distributed GTPase dynamics in the full coupled model}
\label{sec:spatial_results}

Here we numerically solve the full PDE model \eqref{eq:FullModelsEqs} with
dynamic cell size, and forces created at the cell ends by local GTPase. We take
into account dilution, and examine both contractile and protrusive GTPase.
Details of the numerical implementation is given in
\cref{sec:numerical_details}.

\subsection{Single GTPase}

\begin{figure}\centering
\includegraphics[width=0.9\textwidth]{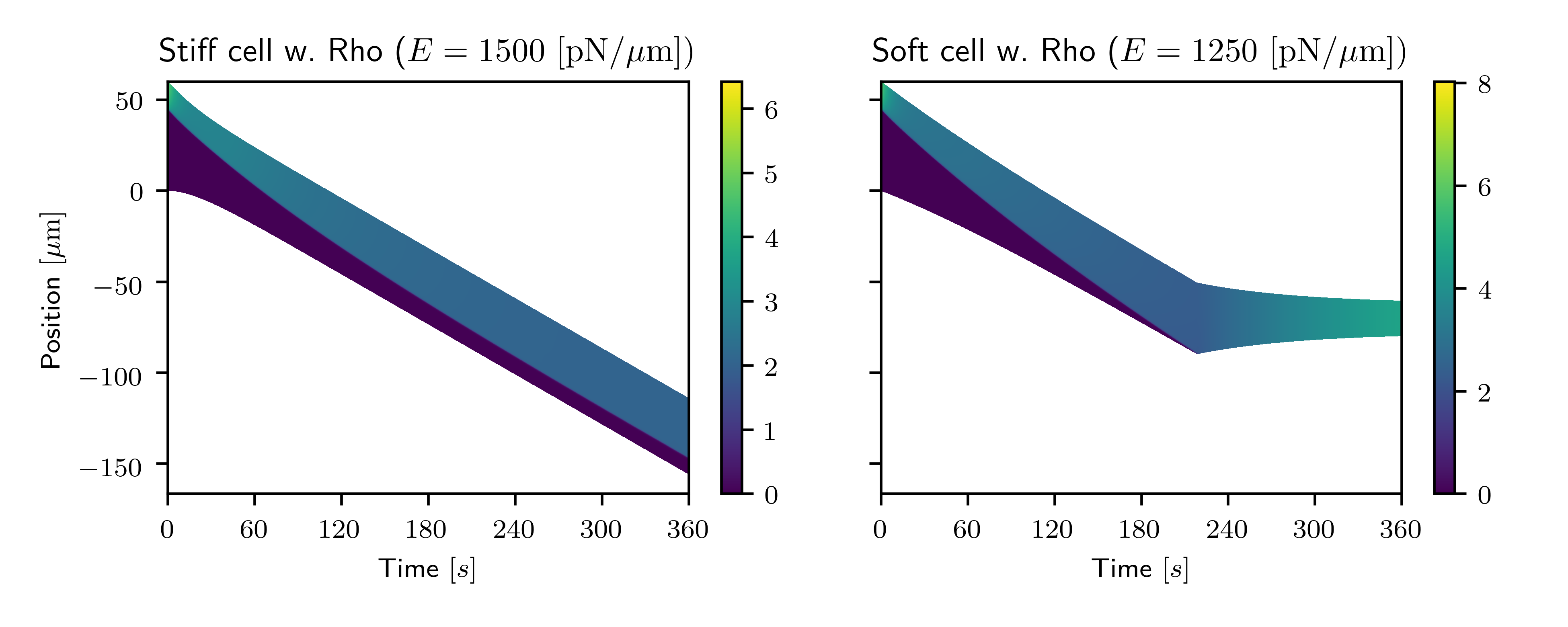}
\includegraphics[width=0.9\textwidth]{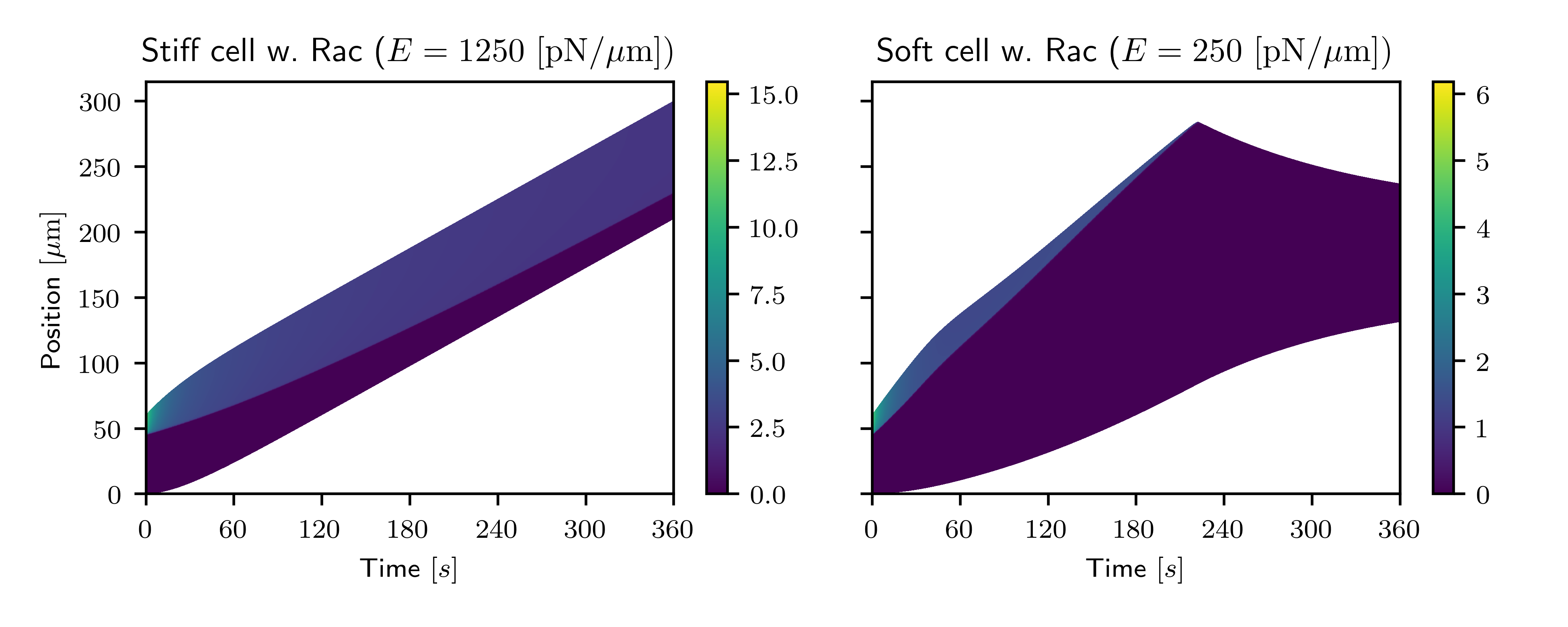}
\caption{Full spatiotemporal simulations of a single GTPase in the 1D cell.
    (Shown is the concentration of active GTPase.) Top: the contractile (Rho)
    case. Bottom: the protrusive (Rac) case.  Initially, cell length is $L(0) =
    L_0$.  Active GTPase initialized with step function at the top edge of the
    cell.  Left: A stiff cell with $G_T = 25$; length hardly changes in either
    case and polarization is maintained.  Right: A soft cell with $G_T = 10$,
    showing significant length change. Once the cell shrinks or expands too
    much, it loses polarization.
}\label{Fig:RhoPol}
\end{figure}

Results are shown as kymographs in \cref{Fig:RhoPol}.  Cells were initialized at
rest-length, $L(0) = L_0$, with elevated GTPase (step function) at one edge. We
tested both stiff (left) and soft (right) cells, by varying the parameter $E$.

For the Rho case (top row of \cref{Fig:RhoPol}), motility is by ``pushing'' or
``squeezing'' from the rear. That is, as the edge of the cell with high Rho
activity contracts inwards, the cell's elastic ``spring'' pushes the other edge
outwards. If the cell is stiff, both edges then translocate in a nearly linear
fashion. A soft cell never manages to compensates for the contraction of the
rear edge.  Eventually, the cell becomes too small to support a polarized GTPase
pattern. Then polarization is lost and the whole cell contracts and keeps
shrinking, amplifying the Rho concentration.

The Rac GTPase case is shown in the bottom row of \cref{Fig:RhoPol}. The cell
moves in the direction of the Rac gradient (edge protrudes outwards). As before,
for stiff cells (left), the rear keeps up with the front due to the restoring
spring-force; a tight GTPase gradient is formed and persists in the moving cell.
In the soft cell (right), the rear edge is only weakly pulled along by the
elastic restoring force. Once the cell is ``too long'', polarization is lost,
Rac activity collapses abruptly to a low uniform level, and the cell shrinks.

From the simulation results (and many replicates, not shown) we draw the
following overall conclusions.  (1) The direction of motion depends on initial
conditions and GTPase type. Cells move towards the direction of Rac and away
from the direction of Rho gradient in the single GTPase simulations.  (2)
Results demonstrate polarization (gain or loss) as well as size changes and the
interplay between these.  (3) Polarized cells can become non-polar by growing or
shrinking too much - both extremes drive the system away from the polarization
parameter regime.  (4) Softer cells are particularly prone to depolarization as
restoring cell size takes longer than the timescale of loss of polarization: see
the bottom row of \cref{Fig:RhoPol}. (Equivalently, cells with large protrusive
or contractile forces $F_{\text{max}}$ are also prone to depolarization.) (5)
Cells with two peaks of Rac at opposite ends tend to stretch and lose the
ability to polarize. Two peaks of Rho cause cells to shrink too much and lose
polarization. (6) Changing the cell's internal viscosity does not strongly
affect the above result, but it changes the time scale of cell deformation.

%%%%%%%%%%%%%%%%%%%%
\subsection{Mutual antagonism: Rac and Rho in the same cell}
\label{sec:RacRho_results}

Finally, we simulated the Rac-Rho mutually antagonistic full model as described
in the ``Two GTPase model'' of \cref{sec:Modeleqs}. The forces of protrusion or
retraction on the cell end $x_k$ are assumed to depend on both Rac and Rho as
follows:
\[
  F_{k}(t) = \lb -1 \rb^{k} F_{\text{max}}
  \lb \frac{R^m(x,t)}{G_c^m + R^m(x,t)} - \frac{\rho^m(x,t)}{G_c^m + \rho^m(x,t)}\rb
  \Big|_{x = x_{k}},
\]
for $k=1,2$. In the full model equations~\eqref{eq:FullModelsEqs}, this replaces
equations~\eqref{eq:cell_forces}. Note that since Rac and Rho are mutually
antagonistic, one will dominate at each of the cell endpoints.

\begin{figure}\centering
    \includegraphics[width=0.95\textwidth]{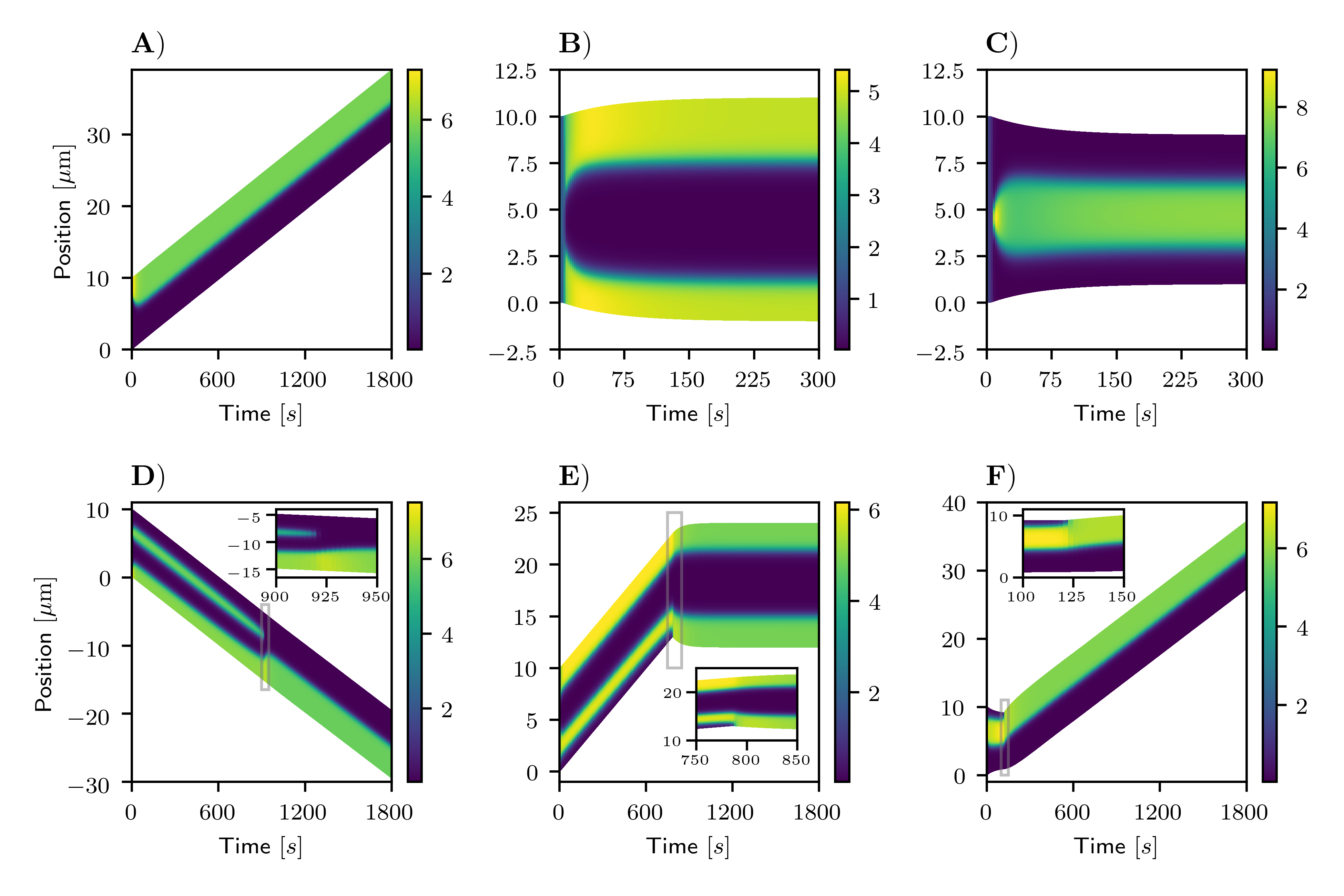}
    \caption{Various possible behaviours in the full Rac-Rho model with mutual
    inhibition. Shown is the profile of active Rac. (Active Rho profile is
    complementary.) Parameters are $L_0 = 10 \si{\micro\metre}, m, n= 4, \gamma
    =5, G_T = 4$.  Plots differ in initial inactive Rho (small Gaussian noise
    about the uniform steady state); other GTPase initially uniform across the
    cell. (A) Polarized cell with Rac front moves at constant rate. (B)
    Stationary cell with high Rac at both ends expands.  (C) Cell with high Rho
    at both ends contracts without moving.  (D) Cell with two peaks of Rac
    becomes uni-polar (at around 900 s) and continues to move in the same
    direction.  (E) Similar to D, but with reorganized Rac eventually stalling
    the cell. (F) Initially static cell with an internal region of high Rac,
    which moves toward a boundary, thus mobilizing the cell.  Insets: expanded
    views of regions of interest shown in grey rectangular portions of the
    trajectories.
    }\label{Fig:SingleCellRacRhoLeft}
\end{figure}

Cells were initialized with the same parameters, spatially uniform GTPase, with
superimposed Gaussian noise in inactive Rho.  Sample results are shown in
\cref{Fig:SingleCellRacRhoLeft}.

These results (including others, not shown) can be summarized as follows: (1)
The evolving GTPase pattern is determined by the initial conditions.  (2) Rac
and Rho segregate to mutually exclusive regions in the cell.  (3) Polarized
motile cells spontaneously gain active Rac at the ``front'' and Rho at the
``rear''.  (4) Cells with more than a single transition layer display rich
dynamics, in particular as plateaus of active GTPase transition from the
interior to a cell edge.  This causes some cells to stall, and others to start
moving, as shown in \cref{Fig:SingleCellRacRhoLeft}.  (5) Cells with high Rac at
both ends get larger, whereas those with high Rho at both ends shrink. Such
metastable states can persist for a long time.

In short, the full model with Rac-Rho mutual inhibition displays a diverse set
of behaviors not seen in model variants that neglect the full spatio-temporal
GTPase distribution, or that assume static cell size.

\section{Discussion}
\label{sec:discussion}

Interplay between the activity of GTPases, the contraction or expansion of a
cell, and resultant feedback to GTPase activation was previously
explored in a related framework by \citep{Zmurchok2018}. Our approach here
differs in two main respects: (1) GTPase is no longer assumed to be spatially
uniform, and (2)  cell size dynamics is considered in the mass balance of the
GTPase.

Accounting for the spatial distribution of the GTPases inside a cell leads to
new features that were absent in \citep{Zmurchok2018}. These include (a) delay
while the internal GTPase activity reorganizes spatially, (b) the possibility of
polarization and net motion, rather than just cell expansion or contraction (c)
cases where the cell stalls when plateaus of GTPase form at both cell edges, and
(d) loss of polarization if the cell gets too large or too small.

Despite the simplicity of the model, initially comprised of a single GTPase
(with active/inactive forms) in 1 spatial dimension, we find a variety of
possible behaviours even in the spatially homogeneous case, depending on the
action of the GTPase. The simplicity of the model then permits analysis of steady
states, stability, and Hopf bifurcations. Analysis is considerably more challenging
in the more detailed models.

There are two ways that cell size affects the results of bifurcation analysis.
1) through the definition of length and time scales; and 2) by diluting or
concentrating the internal signaling activity, as shown by detailed mass
balance, and known classically. This in turn, also affects whether the cell can
have a polarized state. The single GTPase model of Eqs.~\eqref{Eqn:General_GTP_model} can sustain
``polarized'' wave-pinning solutions, but if the domain grows (or shrinks) too
much these are lost. This is consistent with \citep{mori2011} where the
bifurcation parameter $\epsilon$ depended on cell size.

We showed that the cell could undergo cycles
of expansion and contraction in the case of a Rac-like GTPase that pushes out
the cell edges.  Recent work on cell oscillations (in the context or confined epithelial sheets) is given in \citep{Peyret2019}, and a review of cycles of protrusion and retraction in experimental
and theoretical models of the cell edge is found in \citep{Ryan2012}. Cycles also
occur in paper by \citep{Dierkes2014}, who consider a similar treatment of
dilution (of myosin concentration) when the cell stretches, and a mechanical
contractile unit driven by myosin. Unlike our model, their chemical equation is
simply linear return to a baseline level $c_0$, with a similar dilution term
$(c/\ell) (d\ell/dt) $. Their oscillations stem from an assumption that the
``spring'' governing cell length has a nonlinear (cubic) term ($K(\ell)$).

One can argue that some cells deform and redistribute their protrusions while
maintaining a fixed volume. For this reason, we also considered a model variant
with fixed volume. Qualitatively, the two variants are similar, through some
features differ. See Appendix for a comparison.  It is easy to generalize the
model \eqref{Eqn:CellKV} to the version in the recent paper by \citep{Bui2019}
who used distinct values of the frictional coefficient $\mu$ at the front and
the rear of each cell.

While this study demonstrated the behaviour of a simple GTPase system in 1D, it
prepares the ground for 2D simulations where cell shape can be explored more
fully. It also opens the way to considering how cells interact with one another and
how mechanochemical feedback between those cells affects emergent behavior at
the scale of a tissue.

While it is straightforward to carry out analysis in the uniform case, such
analysis is more challenging in the spatially distributed cases. Since in our
model the forces on the cell ends solely depend on the active GTPase
concentrations at the cells ends, a worth while question for future research
is whether it is possible to replace (in some limit) the full PDE by two ODEs
tracking the amount of active GTPase at the cell ends.
A distinct challenge is the emergence of more complicated transient dynamics
(see \cref{Fig:SingleCellRacRhoLeft}), which persist on time-scales significant
for cellular behaviour. This highlights the need to not only better understand
the asymptotic behaviour of such equations but also their transient dynamics.
This becomes even more urgent in view of future models containing more details
and multiple cells.

%%%%%%%%%%%%%%%%%%%%%%%%%%%
%%%%%%%%%%%%%%%%%%%%%%%%%%%
\clearpage
%%%%%%%%%%%%%%%%%%%%%%%%%%%%%
%%%%%%%%%%%%%%%%%%%%%%%%%%%%%%

\appendix
\section{Mass balance on growing or shrinking domain}
Consider a prototypical reaction diffusion equation for some concentration
$u(x,t)$ in conservative form,
\[
 u_t = - \nabla\cdot\mathbf{j}(u) + f(u), \quad\mbox{in}\ \Gamma(t),
\]
where $\mathbf{j}(u)$ is the particle flux. Further, assume no-flux boundary
conditions on the boundary of the domain, $\partial \Gamma(t)$. In integral
form, we write a conservation law for a small volume element $V(t) = [x(t), x(t)
+ \Delta x(t)]$
\[
    \frac{\dd}{\dd t} \int_{V(t)} u(x, t) \dd x
      = \int_{V(t)} \lsb -\nabla\cdot\mathbf{j} + f(u) \rsb \dd x.
\]
Using the Reynolds transport theorem the left hand side is equivalent to
\[
    \frac{\dd}{\dd t} \int_{V(t)} u(x, t) \dd x
    = \int_{V(t)} \lsb u_t + \nabla\cdot\lb\mathbf{v} u\rb\rsb \dd x,
\]
where $\mathbf{v}$ is the velocity of the material point $x(t)$.
We then obtain the following reaction-transport-diffusion equation
\begin{equation}\label{Eqn:App_RD_Var_Domain}
  u_t + \nabla \cdot \lb \mathbf{v} u \rb = D \Delta u + f(u), \qquad\mbox{in}\ [x_1(t), x_2(t)].
\end{equation}
Similarly, the boundary conditions on $\Gamma(t)$ become
\[
  -D \nabla u + \mathbf{v} u = 0, \quad\mbox{on}\ \partial\Gamma(t).
\]
We have two new terms
\begin{enumerate}
    \item An advection term, $\mathbf{v} \cdot \nabla u$, which corresponds to
      elemental volumes moving with the flow due to local growth.
    \item A dilution term, $u \nabla\cdot \mathbf{v}$, due to local volume changes.
\end{enumerate}

\section{Numerical Implementation}\label{sec:numerical_details}

\subsection{Mapping to a constant domain}

It is most convenient to discretize and numerically solve the PDEs on a constant
domain. Hence a mapping is used to transform the growing/shrinking cell to the
interval $[-1,1]$.  This can be achieved by the following change of variables
\[
 (x,t) \mapsto (\bar x, \bar t) = \left(\frac{x - x_c(t)}{R(t)}, t \right).
\]
Note that $\bar{x}$ is the location in the new domain $[-1,1]$,
whereas $x$ is located in the cell domain $[x_1(t), x_2(t)]$.
Carefully applying the chain-rule leads to
\[
\begin{split}
 u_{\bar{x}} &= \frac{\partial u}{\partial x} \frac{\partial x}{\partial {\bar x}}=
 u_x R(t) \\
 u_{\bar{t}} &= \frac{\partial u}{\partial t} \frac{\partial t}{\partial {\bar t}}+ \frac{\partial u}{\partial x} \frac{\partial x}{\partial t}\frac{\partial t}{\partial {\bar t}}
 = u_t + u_x \lb \bar{x} R'(t) + x_c'(t) \rb,
\end{split}
\]
where we have used $x={\bar x}R(t)+x_c(t)$ and $t={\bar t}$ in the transformation.
Substituting these results into Eqn.~\eqref{Eqn:RD_Var_Domain}, we obtain

\[
 u_{\bar t} - u_{\bar x} \frac{\bar{x} R'(t) + x_c'(t)}{R(t)}
 + \frac{x_c'(t) + \bar{x} R'(t)}{R(t)} u_{\bar x}
 + u \frac{R'(t)}{R(t)} = \frac{D}{R^2(t)} + f(u), \ \mbox{in}\ [-1, 1].
\]

Cancelling two terms on the LHS. and rearranging leads to
\[
    u_{\bar t} = \frac{D}{R^2(t)}u_{\bar{x}\bar{x}} + f(u) - u \lb\frac{\dot
    R(t)}{R(t)}\rb,\ \mbox{in}\ [-1,1].
\]

The total amount of GTPase on the interval $[-1,1]$ becomes time dependent
through
\[
 G_T = \int_{-R(t)}^{R(t)} \lb G + G_i \rb \dd x = R(t) \int_{-1}^{1} \lb G +
 G_i \rb \dd y.
\]

%%%%%%%%%%%%%%%
\subsection{Changing rest length is equivalent to imposing additional force}

It is easy to see, from Eqn.~\eqref{eq:dLi/dt} that adjusting the cell's rest
length, $L_0$ is equivalent to assuming that additional forces act on the cell:
\begin{equation}
  (\mu+2\eta) \frac{\dd L}{\dd t} = (F_2-F_1 ) -2 E(L - L_0) = -2 E(L - (L_0 + \Delta L_0)),
\label{eq:dLi/dt2}
\end{equation}
where $\Delta  L_0 = (F_2 - F_1) / 2E$. In other words a change of $\Delta L_0$
is equivalent to an additional force of $-2E\Delta L_0$.

In \citep{Zmurchok2018} and \citep{Liu2019} only the spatially
uniform $G$ model variants were considered. There, cell length was modeled by \eqref{eq:dLi/dt2}, with
\[
  \Delta L_0 (G)= \pm b \frac{G^m}{G_c^m + G^m}.
\]
Rac (+) is assumed to increase $L_0$ and Rho (-) to decrease it, so the sign of
the Hill function depends on the given GTPase.

As shown here, the assumptions that GTPases work through modification of the
cell's effective rest length is mathematically equivalent to assumptions we make
about GTPase-dependent load forces on the cell ends.

%%%%%%%%%%%%%%%%
\subsection{Numerical methods}

Each cell introduces a set of mechanical equations for its endpoints i.e.\
equation~\eqref{Eqn:MechanicalFinal}. Next, each cell contains $K$ GTPases, that
is we include $K$ sets of reaction-diffusion
equations~\eqref{Eqn:General_GTP_model}.  The cell interior is divided into a
cell-centered grid with uniform length $h = 1 / N$, where $N$ is the number of
grid cells per unit length. Here we use $N = 2^{10}$. The second order
derivative is discretized using a second order cell centered finite difference.
For more detailed information on the numerical method, see
\citep{Gerisch2001}. This results in $M$ total grid points per GTPase species, and $K$
GTPase species. The resulting state vector is given by
\[
    y =
    \begin{pmatrix}
        x_1 \\
        x_2 \\
        G_{1, 1} \\
        G_{1, 2} \\
        \vdots \\
        G_{K, M}
    \end{pmatrix}
\]
where $G_{j, k}$ denotes the concentration of the $k$-th GTPase at the $j$-th spatial
grid point. The temporal evolution of the state of the cell can be written as:
\[
    \frac{\dd y}{\dd t} = f(t, y),
\]
where $f(t, y)$ is a function whose first
two entries contain equation~\eqref{Eqn:MechanicalFinal} and the final $KM$
entries are the result of the spatial discretization of the GTPase PDEs.

The resulting system of ordinary differential equations are integrated using the
ROWMAP integrator \citep{Weiner1997a}, an specialized integrator for large
stiff ODEs (such systems are commonly the result of spatially discretizing PDEs).
Here we use the ROWMAP implementation provided by the authors of
\citep{Weiner1997a}\footnote{\url{http://www.mathematik.uni-halle.de/wissenschaftliches_rechnen/forschung/software/}}.
The ROWMAP-integrator (written in Fortran) is wrapped for easy use in a
\texttt{Python} environment using
\texttt{f2py}\footnote{\url{https://docs.scipy.org/doc/numpy-dev/f2py/}}
(f2py provides a connection between the Python and Fortran languages)m
and integrated into \texttt{scipy}'s\footnote{\url{https://www.scipy.org/}}
integrator class (a package providing several different techniques to integrate
ODEs). The spatial discretization (right hand side of ODE) is implemented using
\texttt{NumPy}\footnote{\url{www.numpy.org}}. The error tolerance of the integrator
is set to $v_{tol} = 10^{-6}$.

%%%%%%%%%%%%%%%%%%%%%%%%
\section{Linear stability of spatially uniform GTPase model}\label{subsec:UniformLinStab}

In this section we consider the steady states and their linear stability of
equation~\eqref{Eq:DynSysUnif}.

The Jacobian of the ODE system \eqref{Eq:DynSysUnif} at any steady state $(L_{eq}, G_{eq})$ is
\[
    J(L_{eq}, G_{eq}) = \begin{pmatrix}
        -\frac{2E}{\mu + 2 \eta} & \pm \frac{2F_{\text{max}}}{\mu+2\eta} \frac{mG_c^m G_{eq}^{m-1}}{\lb G_c^m + G_{eq}^m\rb^2} \\
        -\frac{G_T}{L_{eq}^2}\lb b + \gamma \frac{G_{eq}^n}{1 + G_{eq}^n}\rb &
        \frac{\gamma n G_{eq}^{n-1}}{\lb G_{eq}^n + 1 \rb^2}\lb \frac{G_T}{L_{eq}} - G_{eq}\rb
        - \lb b + I + \gamma \frac{G_{eq}^n}{1 + G_{eq}^n}\rb
    \end{pmatrix}.
\]
Model parameters are all positive, and $L_{eq}, G_{eq} > 0$ so in the Rac and
Rho cases, respectively, the Jacobian has the sign patterns:
\[
    J_{\text{Rac}} = \begin{pmatrix} - & + \\ - & \sgn(d) \end{pmatrix}, \qquad
    J_{\text{Rho}} = \begin{pmatrix} - & - \\ - & \sgn(d) \end{pmatrix}.
\]
where
\[
    d \coloneqq \frac{\gamma n G_{eq}^{n-1}}{\lb G_{eq}^n + 1 \rb^2}\lb \frac{G_T}{L_{eq}} - G_{eq}\rb - \lb b + I + \gamma \frac{G_{eq}^n}{1 + G_{eq}^n}\rb.
\]
Note that since $(L, G) \in \Delta$, where $\Delta$ is the invariant region in
\cref{Lemma:InvariantRegion}, we have that $d$ is the difference of two
positive numbers.
It is straightforward to conclude the following (see for instance \citep{murray1989}). In
the Rho case, the steady states are either stable nodes or saddles and no oscillations
are possible. In the Rac case the steady states are stable when $\sgn(d) <
0$, and oscillations are possible when $\sgn(d) > 0$.

\subsection{Sharp-switch limit}\label{subsec:SharpSwitch}

Using a sharp-switch approximation we can easily compute two steady states:
\begin{enumerate}
    \item
When both nonlinear switches are off, i.e.\ $G \ll 1$, the steady states are:
\[
    L_{eq} = L_0, \qquad G_{1} = \frac{b G_T}{L_0 (b + I)}.
\]
Since the linearization of \eqref{Eq:DynSysUnif}
is lower-triangular, the eigenvalues are readily found
\[
    \lambda_1 = \frac{-2E}{\mu + 2\eta}, \quad \lambda_2 = -(b + I).
\]
Since $b, I, E, \mu, \eta >0$, we conclude that this steady state is a stable node. Finally,
since $G_{eq} \ll 1$, we obtain an existence condition
\[
    G_T < L_0 \lb 1 + \frac{I}{b} \rb.
\]
Therefore, the steady state of a large cell, with low total active GTPase can only exist
if the total amount of GTPase is small.

\item In the case $G_c < G < 1$, i.e.\ one of the nonlinear switches is turned
on, the steady states are:
\[
    L_{eq} = L_0 \pm \frac{F_{\text{max}}}{E},
        \qquad G_{2} = \frac{b G_T}{L_{eq}(b + I)}.
\]
Since the linearization of \eqref{Eq:DynSysUnif} is lower-triangular as before, the eigenvalues
are readily found
\[
    \lambda_1 = \frac{-2E}{\mu + 2\eta}, \quad
    \lambda_2 = -(b + \gamma + I).
\]
Since $b, I, \gamma, E, \mu, \eta >0$, we conclude that this steady state
is a stable node. A condition for the existence of this steady state is
\[
    \frac{L_0 \lb b + \gamma + I\rb}{b + \gamma} < G_T < \frac{G_c L_0 \lb b + \gamma + I \rb}{b + \gamma}.
\]
Thus we must have some intermediate amount of GTPase.

\item In the case $1 < G < G_c$, i.e.\ one of the nonlinear switches is turned
on, the steady states are:
\[
    L_{eq} = L_0, \qquad G_{2} = \frac{(b + \gamma) G_T}{L_0(b + \gamma + I)}
\]
Eigenvalues
are readily found, as before,
\[
    \lambda_1 = \frac{-2E}{\mu + 2\eta}, \quad
    \lambda_2 = -(b + \gamma + I).
\]
From $b, I, \gamma, E, \mu, \eta >0$ we conclude that this steady state
is a stable node. The existence condition for this steady state can be written
in either of two forms
\[
 \frac{L_0 \lb b + \gamma + I\rb}{b + \gamma} < G_T < \frac{G_c L_0 \lb b + \gamma + I \rb}{b + \gamma},
 \quad
 \frac{L_0}{G_T-L_0}I-b < \gamma < \frac{L_0 G_c}{G_T-L_0 G_c}I-b.
\]
For these ranges to be non-empty we require that $G_T > L_0 G_c$. Thus we must
have some intermediate amount of GTPase.

\item
In the case, when both nonlinear switches are turned on, i.e.\ $G \gg G_c$, the steady states are
\[
    L_{eq} = L_0 \pm \frac{F_M}{E}, \qquad G_{3} = \frac{(b + \gamma) G_T}{L_{eq}(b + \gamma + I)}.
\]
Then the eigenvalues are
\[
    \lambda_1 = \frac{-2E}{\mu + 2\eta}, \quad
    \lambda_2 = -(b + \gamma + I).
\]
Since $b, I, \gamma, E, \mu, \eta >0$, this steady state
is a stable node. The condition for existence of this steady state is given by
\[
 G_T > G_c L_{eq} \lb1 + \frac{I}{b + \gamma}\rb, \ \mbox{or}\ \
    \gamma > \frac{L_b G_c}{G_T-L_b G_c}I-b.
\]
Thus only cells with a sufficiently amount of GTPase can be large or small.
\end{enumerate}

In the bifurcation diagram of \cref{Fig:BothBifurcYLiu}, the regions of existence of $G_1,
G_2$ or $G_3$ are identified.

\section{The case of a deforming cell with conserved volume}
\label{sec:const_vol}

Here we consider the case that
the volume of a cell is roughly constant, while the cell deforms and its membrane size changed. This idea suggests a slight modification to our model.
The modified mass conservation reads
\[
 G_T = G_i V + \beta G L,
\]
where $V$ is the constant volume of the cell, and $\beta$ is a constant of
proportionality with units of [length]$^2$. We can solve for $G_i$ to get a
relation analogous to \eqref{eqn:mass_conserv}:
\[
 G_i = \frac{G_T}{V}-\frac{\beta GL}{V}.
\]
Notice that after substituting this into the well-mixed system, we can set $V=1$
by rescaling $b$ and $\gamma$. Therefore, we get
\begin{equation}\label{eqn:constvol_model}
 \frac{\dd G}{\dd t} = \lb b + \gamma \frac{G^n}{1+G^n}\rb\lb G_T - \beta G L\rb - IG - G\lb \frac{\dot{L}}{L}\rb.
\end{equation}
This, together with \eqref{Eq:dL/dt_uniform_G} and \eqref{eq:Force_uniform_G},
forms our new system.

Much of the analysis of this system is similar to the earlier case. The changes
to the nullclines in the $G-L$ phase plane correspond to a vertical shift in the
sharp switch limit, which can be compensated by adjusting the parameters. This
suggests that the behaviors of the new system is similar to the original
\eqref{sys:uniform_G}.  The equilibrium branches and the conditions
for their existence in the sharp switch limit are:
\begin{enumerate}
    \item For $G \ll 1$:
    \[L_{eq}=L_0, \quad G_{eq}=\frac{b G_T}{b \beta L_0+ I}
    \quad \mbox{exists provided} \quad
    G_T < \beta L_0 + \frac{I}{b}.
    \]
    \item For $G_c<G<1$:
    \[L_{eq}=L_0\pm \frac{F_{max}}{E}, G_{eq}=\frac{b G_T}{b \beta L_{eq}+ I}
     \quad \mbox{exists provided} \quad
    G_c \lb \beta L_{eq}+\frac{I}{b} \rb<G_T <\beta L_{eq}+\frac{I}{b}.
    \]
    \item For $1<G<G_c$:
    \[L_{eq}=L_0, G_{eq}=\frac{(b+\gamma) _T}{(b+\gamma) \beta L_0+ I} \quad \mbox{exists provided} \quad
    \beta L_0 + \frac{I}{b+\gamma}< G_T < G_c \lb \beta L_0 + \frac{I}{b+\gamma} \rb.\]
    \item For $G \gg G_c$:
    \[L_{eq}=L_0 \pm \frac{F_{max}}{E}, G_{eq}=\frac{(b+\gamma) G_T}{(b+\gamma) \beta L_{eq}+ I}  \quad \mbox{exists provided} \quad
    G_T > G_c \lb \beta L_{eq} + \frac{I}{b+\gamma} \rb.\]
\end{enumerate}

\begin{figure}
    \centering
    \includegraphics{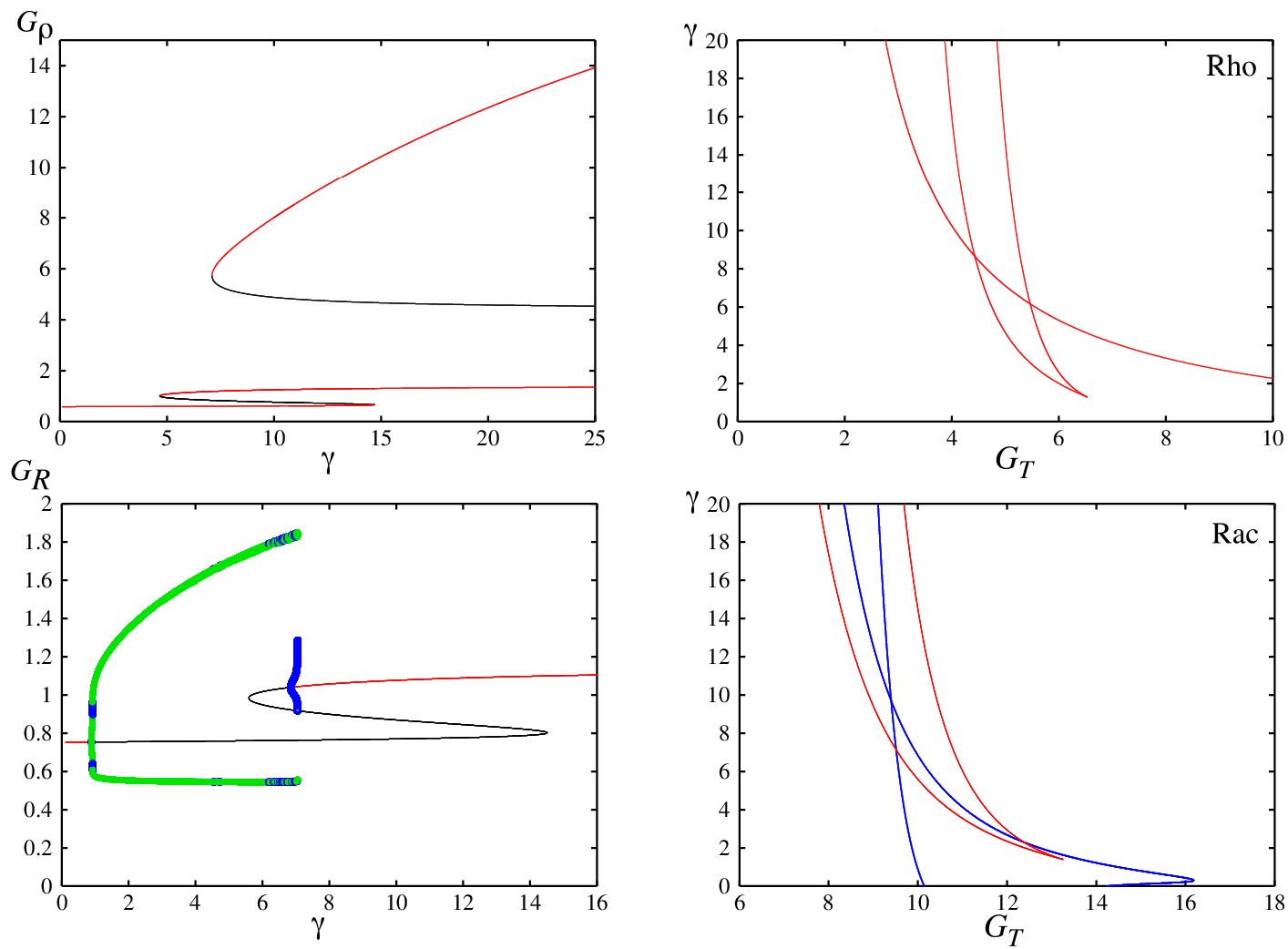}
    \caption{Bifurcation diagrams as in  \cref{Fig:BothBifurcYLiu}, but for the constant volume spatially uniform GTPase model, \eqref{eqn:constvol_model}. Top: the contractile (Rho) case,
    with parameters
    $k=1,I=5,L_0=3.5,\frac{2E}{\mu+2\eta}=3,\frac{2F_{max}}{\mu+2\eta}=10,G_c=4,\beta=1,G_T=5,m=n=10$.
    Bottom: the protrusive (Rac) case, with parameters
    $k=1,I=10,L_0=2.3,\frac{2E}{\mu+2\eta}=8,\frac{2F_{max}}{\mu+2\eta}=55,G_c=0.9,\beta=1,G_T=10,n=20,m=10$.
    While the Rac case is still tristable, a fold point that previously connected the middle and high branch (in the orignal model) is now
    missing. These branches persists as $\gamma \to \infty$ whereas the lower
    branch exist only for low $\gamma$. In the Rac case, the Hopf curve in the
    two-parameter bifurcation digram (lower right) differs significantly from the corresponding panel in
    \cref{Fig:BothBifurcYLiu}, which yield parameter regimes
    where the limit cycle exists for $\gamma \to 0$.} \label{fig:constvol_bif}
\end{figure}

In the Rho (contraction) case, tri-stability still occurs. In this model, we can
more readily find parameter sets where the middle ($1<G<G_c$) and high
branch persist as $\gamma \to \infty$, whereas the low branch terminates at a
fold point (\cref{fig:constvol_bif}). (This is also possible in the
original model.) This suggests that even with a very large positive feedback in
GTPase activity, and when GTPase activity is high enough to trigger that
feedback, the cell can still stay at a large size. In contrast, in the original
model, the regime where only the middle and high branch coexist is usually
bounded above by some maximum $\gamma$, as seen in the regime labelled ``2,3" in \cref{Fig:BothBifurcYLiu} (top right).

In the Rac (expansion) case, periodic orbits are still possible. The regime in which
limit cycles exist is much wider that in the original model. We can identify
parameter sets where a limit cycle exists even at $\gamma=0$, which hints that in
this model, a positive feedback in GTPase activity is not neccessary for an
oscillating cell.

Overall, the model variant with constant cell volume has the same
regimes as the original model. However, the location and size of these regimes
has shifted.

The Jacobian of the ODE system at a steady state
$(L_{eq}, G_{eq})$, is
\[
    J(L_{eq}, G_{eq}) = \begin{pmatrix}
        -\frac{2E}{\mu + 2 \eta} & \pm \frac{2F_{\text{max}}}{\mu+2\eta} \frac{mG_c^m G_{eq}^{m-1}}{\lb G_c^m + G_{eq}^m\rb^2} \\
        -\frac{2E}{\mu + 2 \eta} \frac{1}{L_{eq}} &
        \frac{\gamma n G_T G^{n-1}}{\lb 1+G^n \rb^2} - \beta L\lb b + \gamma \frac{G^n}{1+G^n} + G \rb - I
    \end{pmatrix}.
\]
Model parameters are all positive, and $L_{eq}, G_{eq} > 0$ so in the Rac and
Rho cases, respectively, the Jacobian has the sign patterns:
\[
    J_{\text{Rac}} = \begin{pmatrix} - & + \\ - & \sgn(d) \end{pmatrix}, \qquad
    J_{\text{Rho}} = \begin{pmatrix} - & - \\ - & \sgn(d) \end{pmatrix}.
\]
where
\[
    d \coloneqq
        \frac{\gamma n G_T G^{n-1}}{\lb 1+G^n \rb^2} - \beta L\lb b + \gamma \frac{G^n}{1+G^n} + G \rb - I.
\]
The sign patterns are the same as in the original model, and thus the same conclusions apply.

% Note that since $(L, G) \in \Delta$ the invariant region from
% \cref{Lemma:InvariantRegion}, we have that $d$ is the difference of two
% positive numbers.
% It is straight forward to conclude that (see for instance \citep{murray1989}) in
% the Rho case the steady states are either stable or saddles and no oscillations
% are possible, while in the Rac case the steady states are stable when $\sgn(d) <
% 0$, and when $\sgn(d) > 0$ oscillations are possible.

\bibliographystyle{spbasic}      % mathematics and physical sciences
\bibliography{bibliography}   % name your BibTeX data base

\label{lastpage}
\end{document}